\documentclass{jfm}

\usepackage{xcolor}
\usepackage{graphicx}
\usepackage{amssymb, amsmath, amsfonts}
\usepackage{bm}
\usepackage{newtxtext} % This is the journal default. Please don't change this
\usepackage{newtxmath}
\usepackage{natbib}
\usepackage[justification=justified,format=plain, width=\textwidth]{caption}
\usepackage{siunitx}
\usepackage{orcidlink}
\usepackage{booktabs}
\usepackage{multirow}
\newcolumntype{M}[1]{>{\centering\arraybackslash}m{#1}}
\usepackage{array}
\usepackage{caption}
\usepackage{hyperref}
\hypersetup{
    colorlinks = true,
    urlcolor   = blue,
    linkcolor  = blue, %Colour of internal links
    citecolor  = blue,
}

\usepackage{setspace}

\newcommand{\RomanNumeralCaps}
\linenumbers

\title{Mean-field synchronization model of turbulent thermoacoustic transitions}

\author{Samarjeet Singh\aff{1}
	\corresp{\email{samarjeet.singh448@gmail.com}},
	Amitesh Roy\aff{1}, Jayesh M. Dhadphale\aff{1}, Swetaprovo Chaudhuri\aff{2}, \and  R. I. Sujith\aff{1}} 

\affiliation{\aff{1}Department of Aerospace engineering, IIT Madras, Chennai 600036, India
	\aff{2}Institute for Aerospace Studies, University of Toronto, North York, ON M3H 5T6, Canada }

\begin{document}
\maketitle

%\setstretch{2}
 
\begin{abstract}
%\textit{\textbf{Punchline --} We propose a mean-field synchronization model of thermoacoustic interactions capable of capturing the bifurcations and their criticalities, and the rich phase dynamics which underlie thermoacoustic transitions in disparate combustion systems.}
Thermoacoustic instabilities observed in turbulent combustion systems have disastrous consequences and are notoriously challenging to model, predict and control. Here, we introduce a mean-field model of thermoacoustic transitions, where the nonlinear flame response is modeled as the amplitude weighted response of an ensemble of phase oscillators constrained to collectively evolve at the rhythm of acoustic fluctuations. Starting from the acoustic wave equation coupled with the phase oscillators, we derive the evolution equations for the amplitude and phase and obtain the limit cycle solution. We show that the model captures abrupt and continuous transition to thermoacoustic instability observed in disparate combustors. We obtain quantitative insights into the model by estimating the model parameters from the experimental data using parameter optimisation. Importantly, our approach provides an explanation of spatiotemporal synchronization and pattern-formation underlying the transition to thermoacoustic instability while encapsulating the statistical properties of desynchronization, chimeras, and global phase synchronization. We further show using the model that continuous and abrupt transitions to limit cycle oscillations in turbulent combustors corresponds to synchronization transitions of \textit{second-order} and \textit{first-order}, respectively, of the phase-field comprising the phase difference of pressure and heat release rate fluctuations. We then rationalise our findings in terms of the frequency distribution of oscillators obtained from experiments. The present formulation provides a highly interpretable model of thermoacoustic transitions: changes in empirical bifurcation parameters which lead to limit cycle oscillations amounts to an increase in the coupling strength of the phase oscillators, promoting global phase synchronization. The generality of the model in capturing different types of transitions and states of pattern-formation highlights the possibility of extending the present model to a broad range of fluid-dynamical phenomena beyond thermoacoustics. 
\end{abstract}


\section{Introduction} \label{Sec1_Introduction}
Thermoacoustic instability arising in turbulent combustion systems such as rockets and gas turbine engines remains a significant hindrance to their development \citep{lieuwen2005combustion, sujith2021thermoacoustic}. Thermoacoustic instability happens due to a positive feedback between turbulent flames and the acoustic modes of the combustor, often leading to disastrous consequences. The transition from a stable operation, characterized by broadband combustion noise, to thermoacoustic instability occurs when the heat release rate fluctuations evolve in-phase with acoustic pressure oscillations \citep{rayleigh1878explanation}, and the total acoustic energy arising through the nonlinear feedback from the flame is greater than the net acoustic losses across the boundary of the combustion chamber \citep{chu1965energy, putnam1971combustion}.  

Combustion noise, or the state of ``stable'' combustion, results from volumetric expansion and convective entropy modes \citep{candel2009flame}. These broadband fluctuations, rooted in turbulence, thus exhibit scale invariance and multifractal behavior \citep{gotoda2011dynamic, nair2014multifractality}. The mechanism of transition from such a low-amplitude chaotic state of combustion noise to a high-amplitude periodic state of thermoacoustic instability is a continuing problem of high practical relevance and intense theoretical interest. Traditionally, the onset of thermoacoustic instability has been considered to be a loss of stability of the fixed point solution of the linearized system. Beyond the critical parameter value, or the Hopf point, a pair of complex conjugate eigenvalues cross the imaginary axis, giving rise to limit cycle oscillations through a Hopf bifurcation \citep{lieuwen2002experimental,strogatz2018nonlinear}. This can happen either through primary supercritical and subcritical bifurcations \citep{lieuwen2002experimental,laera2017flame}, or through a secondary bifurcation of an initially stable, primary limit cycle oscillations \citep{ananthkrishnan1998application, roy2021flame, wang2021multi}. The exact nature of bifurcation depends on the nature and order of nonlinearity present in the system \citep{kuehn2021universal}. 

Supercritical Hopf bifurcation is realized when a change in the value of the control parameter beyond the Hopf point leads to a gradual increase in the amplitude of limit cycle oscillations. In contrast, if the system abruptly jumps from a stable fixed point to a limit cycle attractor of large amplitude, the bifurcation is referred to as subcritical Hopf bifurcation. To revert to the stable solution, the control parameter must be reversed past the Hopf point till the fold point is reached. Thus, subcritical transitions are associated with hysteresis and bistability in the solution \citep{strogatz2018nonlinear}. In exceptional settings, however, higher-order nonlinearities in the system can destabilize the stable branch of the limit cycle solution generated through a primary Hopf bifurcation, leading to a secondary fold bifurcation to a high-amplitude limit cycle \citep{ananthkrishnan1998application}. \cite{ananthkrishnan2005reduced} theoretically proposed the possibility of secondary bifurcation to thermoacoustic instability. Since then, secondary bifurcation has been reported in various experimental studies such as in laminar \citep{mukherjee2015nonlinear} and turbulent \citep{roy2021flame, singh2021intermittency, wang2021multi, bhavi2022abrupt} combustion systems.

In addition to the above-mentioned scenarios, the transition from combustion noise to thermoacoustic instability has often been reported to occur through the state of intermittency \citep{gotoda2011dynamic,nair2014intermittency,kheirkhah2017dynamics}. Intermittency refers to a dynamically stable state consisting of low-amplitude chaotic fluctuations randomly interspersed with high-amplitude periodic fluctuations. Further variation in the control parameter results in more frequent bursts of periodic oscillations, manifesting in a continuous ``sigmoid'' type transition diagram. This gradual emergence of periodic dynamics is associated with the emergence of order or coherence in the spatial dynamics \citep{mondal2017onset, george2018pattern}. Thus, the differences in the transition mechanisms make evident the challenge in the development of thermoacoustic models capable of explaining all these transition scenarios.

One of the most successful approaches in modeling thermoacoustic systems is through the measurement of flame response decoupled from the acoustic analysis of the combustor. This approach, when expressed in the frequency domain, is referred to as a flame transfer and describing functions (FTF/FDF) \cite[][]{merk1957analysis,merk1958analysis,schuller2020dynamics}, and when obtained in the time domain, leads to the concept of impulse response functions \citep{polifke2020modeling}. In FTF/FDF modeling, the flame response is illustrated using Bode plots discerning the gain and phase of frequency response. This is complemented by a linear stability analysis of the acoustic network in the frequency domain that reveals the growth rate of eigenmodes in terms of complex eigenfrequencies \citep{noiray2008unified, schuller2020dynamics}. The measurement of the flame impulse response in the method of time delays, on the other hand, allows for a straightforward interpretation of the convective processes underlying the thermoacoustic system. When a transition to limit cycle oscillations is concerned, a bifurcation analysis is usually performed by modeling the flame response as a nonlinear oscillator and considering its feedback on the acoustic fluctuations \citep{dowling1995calculation,nicoud2007acoustic, balasubramanian2008thermoacoustic,subramanian2013subcritical,agharkar2013thermoacoustic, noiray2011investigation,ghirardo2013azimuthal, nair2015reduced}. Insights gained from FTF/FDF and time delay approaches are often utilized for estimating the dynamical flame models utilized in bifurcation studies \citep{noiray2013deterministic, laera2017flame, noiray2017linear, bonciolini2021low}. 

Although these modeling approaches have been useful in revealing the nature of bifurcation and the resulting stability characteristics of the limit cycle, they remain specific to the nature of nonlinearity encoded in the assumed flame model. The explanation for other types of bifurcation in these models required the inclusion of additional higher-order nonlinear terms with little physical justification and sometimes scarce experimental support. Another fundamental drawback of these models is the assumption of a lumped system where the nonlinear contributions of a spatially-extended convective-diffusive-reaction system underlying the premixed flame are parameterized as a temporal--quadratic, cubic, quintic, etc.--function of acoustic perturbations \citep{laera2017flame, noiray2017linear, bhavi2022abrupt}. Consequently, these models cannot explain the rich spatiotemporal synchronization developing concomitantly during the transition to limit cycles \citep{mondal2017onset, pawar2019temporal, guan2019chaos, roy2021flame, singh2021intermittency}. Finally, there has been no clear resolution on what causes thermoacoustic transitions to be abrupt or continuous and has necessitated disparate modeling approaches.

In the present work, we consider a general model where the nonlinear response of the turbulent flame is assumed to comprise an ensemble of non-identical phase oscillators \citep{dutta2019investigating}. These phase oscillators are coupled to the collective rhythm of the thermoacoustic system weighted by the amplitude of acoustic fluctuations. Thus, we assume that each of the phase oscillators evolves according to \citep{strogatz2005crowd}:
\begin{equation}
\frac{d\theta_i}{d\tilde{t}} = \tilde{\omega}_i + \tilde{K}\hat{R}\sin\left[\Phi(\tilde{t})-\theta_i(\tilde{t})\right], \enspace i =1,...,N,
\label{Eq-1.1}
\end{equation}
where $\theta_i$ and $\tilde{\omega}_i$ are the phase and angular frequency of the $i^{\textrm{th}}$ phase oscillator and $\tilde{K}$ describes the coupling strength among the oscillators. The heat release rate fluctuations arising from these phase oscillators can be determined as $\dot{\tilde{q}}^\prime = \tilde{q}_0 \sum_i \sin[\tilde{\Omega}_0\tilde{t}+\theta_i(\tilde{t})]$. The thermoacoustic model is completed by introducing a feedback coupling between the ensemble of phase oscillators and the acoustic pressure having a frequency $\tilde{\Omega}_0$ during limit cycle oscillations and time-varying normalized amplitude $\hat{R}(\tilde{t})$ and phase $\Phi(\tilde{t})$ of the acoustic fluctuations. The acoustic pressure oscillations are governed by the wave equation where the source term comprises the heat release rate fluctuations, $\dot{\tilde{q}}^\prime$. The phase of local heat release oscillations corresponding to individual phase oscillators is, thus, directly coupled to the acoustic fluctuations, while the phase oscillators are equally weighted and indirectly coupled to each other through the acoustic variables. Since all the oscillators are uniformly affected by coupling, the approach is referred to as the \textit{mean-field} approximation \citep{strogatz2000kuramoto}.  

%\begin{figure}
%\centering
%\includegraphics[width=\textwidth]{Mondal.png}
%\caption{Thermoacoustic transition as an emergence of order amidst disorder. Instantaneous phasor field depicting the relative phase between pressure and heat release rate fluctuations inside a turbulent combustor during the state of (a) combustion noise, (b) intermittency and (c) thermoacoustic instability. Reproduced from \cite{mondal2017onset} with permission from Cambridge University Press.}
%\label{fig:mondal_chimera}
%\end{figure}

\begin{figure}
\centering
\includegraphics[width=\textwidth]{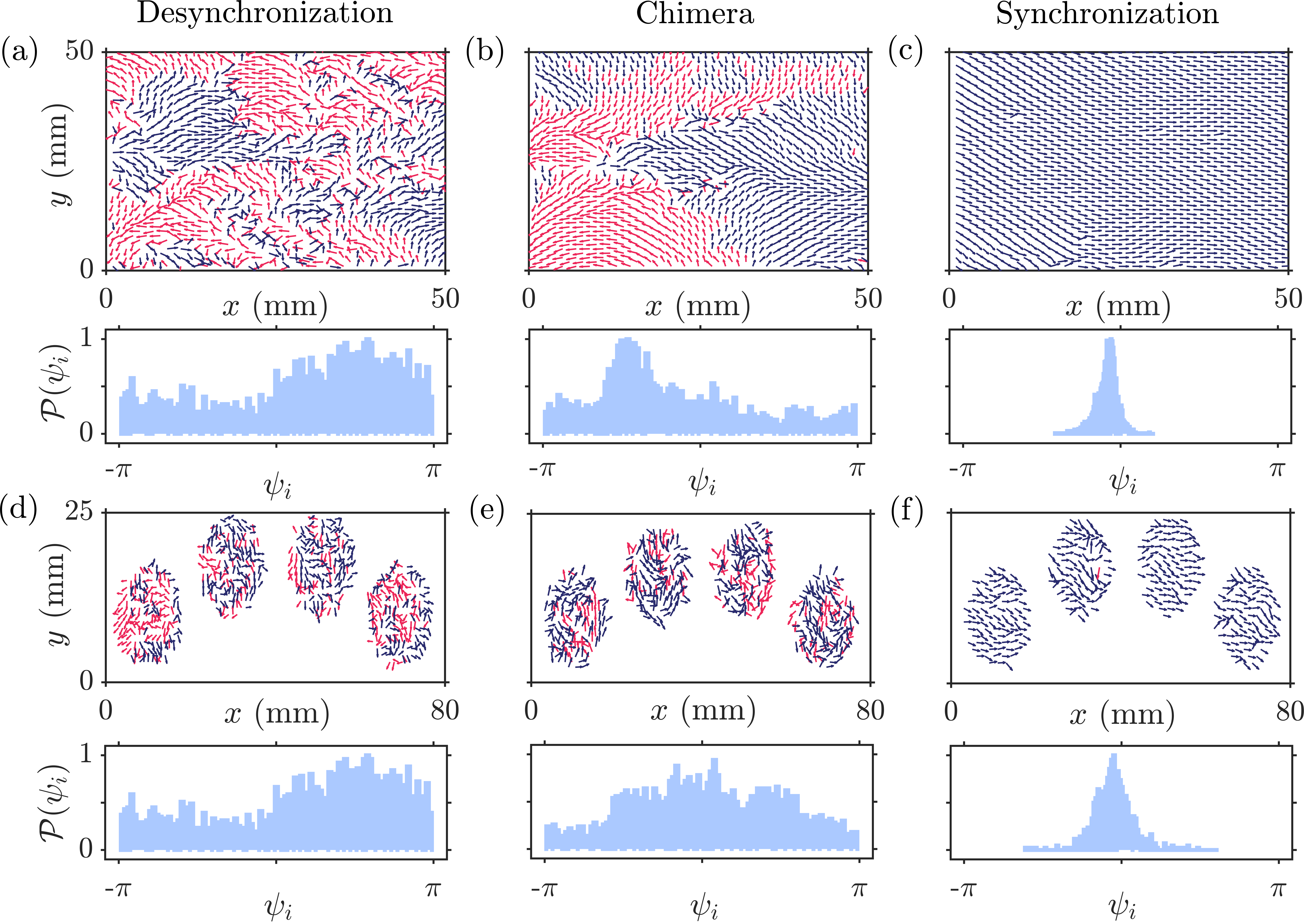}
\caption{Emergence of global synchronization during disparate thermoacoustic transitions. Instantaneous relative phase ($\psi_i$) between the phase of heat release rate fluctuations ($\theta_i$) and acoustic pressure ($\Phi$) is depicted as a field of phasors for the bluff-body (top row) and annular combustor (bottom row) during the state of (a, d) combustion noise, (b, e) intermittency and (c, f) thermoacoustic instability. Phasors have been colored as blue if $|\psi|<\pi/2$ (Rayleigh criteria) and red otherwise to delineate local acoustic power sources from sinks. The probability distribution of the relative phase $\mathcal{P}(\psi_i)$ for each state has also been shown to highlight the appearance of global phase synchronization during thermoacoustic instability. The experimental conditions for the bluff-body and annular combustor correspond to the states shown in figure \ref{fig:Fig5_continuous} and figure \ref{fig:Fig7_seco_timeseries}, respectively. See also supplemental movies S1-S8.}
\label{fig:Fig1_chimera}
\end{figure}

The motivation of the present model comes from the observation that the transition to the state of thermoacoustic instability is associated with the emergence of global phase synchronization of heat release rate and acoustic pressure fluctuations \citep{mondal2017onset,hashimoto2019spatiotemporal,guan2019chaos}. Such an emergence has subsequently been reported in other thermoacoustic systems as well \citep{pawar2019temporal, roy2021flame}. Such a synchronization have been modelled using coupled oscillators \citep{godavarthi2020synchronization, weng2022synchronization} and phase-oscillators \citep{dutta2019investigating, singh2022mean, ramanan2022dynamical}. 

\begin{figure}
\centering
\includegraphics[width=\textwidth]{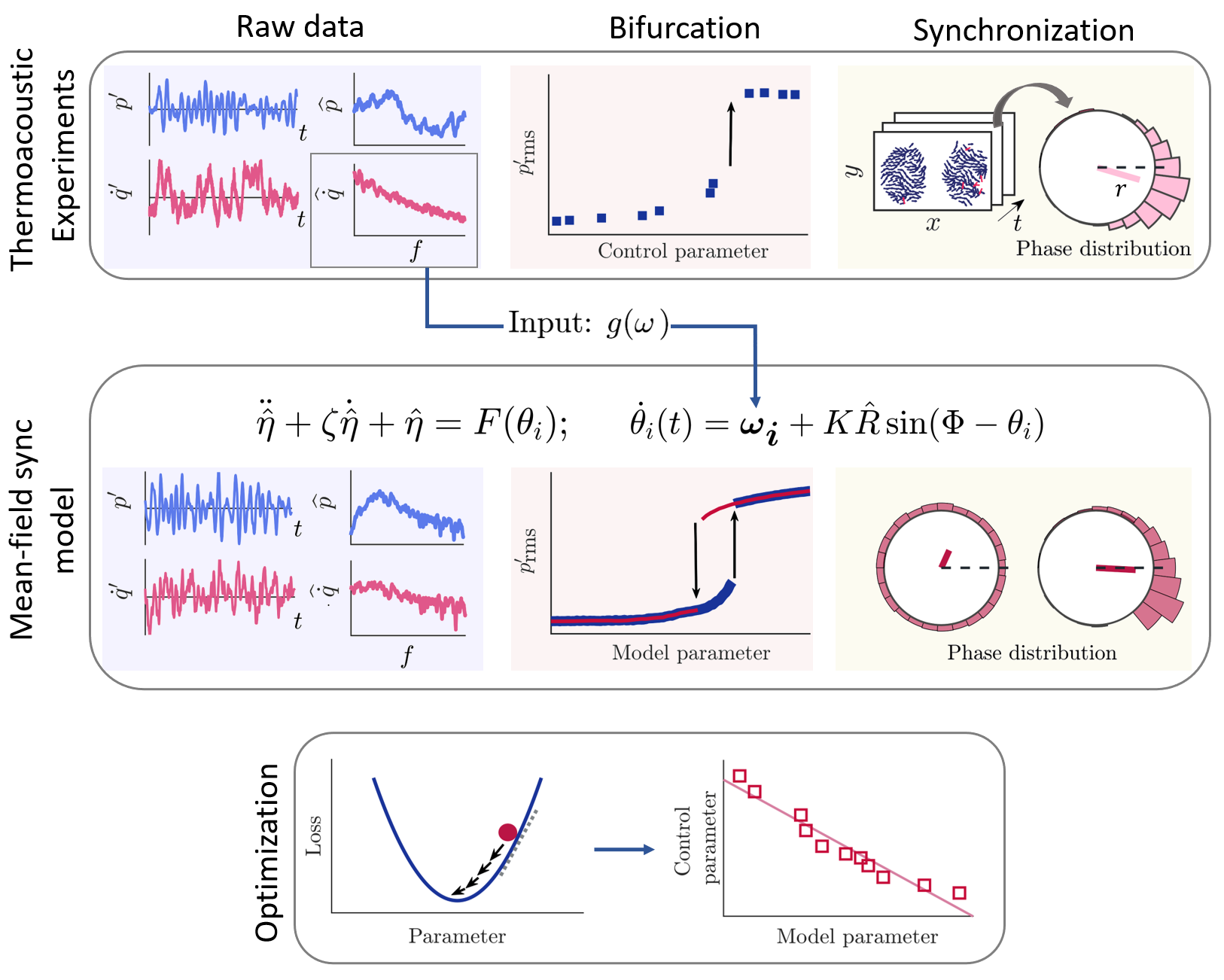}
\caption{Overview of the mean-field model for modeling experimentally observed bifurcation and synchronization. The heat release rate spectrum observed during the state of combustion noise is the only input required for estimating the bifurcation. Parameter optimisation using gradient descent is performed to relate model parameter to the experimental control parameter.}
\label{fig:Fig2_overview}
\end{figure}

Figure \ref{fig:Fig1_chimera} shows the emergence of spatiotemporal synchronization in the bluff-body and annular combustor during the transition to the state of thermoacoustic instability. However, for modeling thermoacoustic transitions along with the underlying states of synchronization, we consider only a dynamic flame model, leaving out any spatial dependencies in view of simplicity. Regardless, one key feature which is incorporated in the model is the nonlinearity in the flame response arising due to effects such as background turbulence, hydrodynamic instabilities, etc. \citep{lieuwen2012unsteady}. These nonlinearities are accounted as weakly coupled interactions among the phase oscillators whose frequencies $\tilde{\omega}_i$ are chosen randomly from a theoretical distribution function $g(\omega)$ or from an experimentally measured heat release rate spectrum. Figure \ref{fig:Fig2_overview} presents the overview, illustrating the manner in which bifurcations and characteristics of synchronization are captured by the model. The form of the mean-field model chosen here is based on its simplicity, analytical tractability, and interpretability. Such models have often been utilized in the study of chemical turbulence and pattern forming fluid dynamical systems \citep{kuramoto1975self, kuramoto2003chemical}, modeling crowd-structure interaction on the Millennium bridge \citep{strogatz2005crowd, abrams2006two, eckhardt2007modeling} and modeling collective interactions in fireflies \citep{ramirez2018fireflies}. 

In this paper, we unveil various aspects of this thermoacoustic mean-field model; we derive the limit cycle solutions for the model using the method of averaging and estimate the amplitude and phase of the limit cycle oscillations. We then depict the applicability of the mean-field model in capturing the transition to limit cycle oscillations in three different types of combustors by taking the heat release rate spectrum during the state of combustion noise as the only input to the model. We show that the model is able to describe both continuous and abrupt transitions to limit cycle oscillations with great accuracy, thus unifying disparate transitions based on the underlying synchronization characteristics. We next implement an optimisation algorithm for model parameter estimation that allows us to pinpoint the exact correspondence between the experimental and the model parameters. Finally, we investigate the features of synchronization in the experiments and the model. We show that continuous and abrupt transitions to the limit cycle oscillations are associated with second-order and first-order synchronization transition in the phase-field describing the phase difference between pressure and heat release rate fluctuations, respectively. We illustrate that even though our model is dynamic with no spatial inputs, it captures the statistical properties of extended spatiotemporal synchronization.

The rest of the paper is organized as follows. In $\S$\ref{Sec2_experimental}, we briefly describe the three different turbulent combustor configurations which show different transition characteristics. In $\S$\ref{Sec3_model}, we expound on the thermoacoustic model. This is followed by an explanation of numerical solution and the optimisation method for estimating model parameters from experimental data in $\S$\ref{Sec4_numer_simu}. We present the results from the model and experiments in $\S$\ref{Sec5_model_pred}. We then discuss the global phase synchronization during the different transitions in $\S$\ref{Sec6_Synchor}. We conclude by summarizing the major findings of our study and discussing their implications in $\S$\ref{Sec7_Conclusion}.

\section{Experimental setup and measurements}
\label{Sec2_experimental}
    
Experiments focusing on the transition to longitudinal thermoacoustic instability were conducted on three disparate turbulent combustors. The combustors have distinct flame holding mechanisms and thermoacoustic characteristics and serve to highlight that the proposed model can be used for predicting transitions and synchronization characteristics under widely different operating conditions (see Table \ref{tab:Expt Features}).

\subsection{Turbulent dump combustor}
Two configurations of the dump combustor were considered: (1) circular bluff-body and (2) fixed-vane swirler stabilized flames. These two configurations show distinct routes to limit cycle oscillations and disparate frequency and amplitude instability. The combustor comprises a plenum connected upstream of a burner tube of 40 mm diameter. The burner tube is connected to the combustion chamber of cross-sectional area $90\times90$ mm (figure \ref{fig:Fig3_experi}a). The burner supports a central shaft of diameter 16 mm, which holds the bluff-body or the swirler in place. Fuel is delivered to the burner tube by four injection holes in the shaft, each of diameter 1.7 mm, 120 mm upstream of the flame holder. Ignition of the combustible air-fuel mixture is facilitated by an $11$ kV spark plug flush mounted on the dump plane. Combustion by-products are exhausted into the atmosphere via an acoustic decoupler of size $1000\times500\times500$ mm. 

\subsubsection{Bluff-body stabilized configuration}
The flame holder is a cylindrical bluff-body (diameter $47$ mm and width $10$ mm) mounted on the central shaft, located 45 mm downstream of the dump plane (figure \ref{fig:Fig3_experi}c). The combustion chamber is $1100$ mm in length. In this configuration, experiments were performed by maintaining a constant rate of fuel flow $\dot{m}_f = 8 \times 10^{-4}$ kgs$^{-1}$ and varying the rate of air flow in steps from $\dot{m}_a = 1.0$ $\times$ $10^{-2}$ to 1.6 $\times$ $10^{-2}$ kgs$^{-1}$. Consequently, the equivalence ratio varies from $\phi=0.86$ to $0.56$ and the Reynold number ($Re$) from $2.2\times 10^4$ to $3.2\times 10^4$. 

A photomultiplier tube (PMT) and pressure transducers are used for measuring pressure and heat release rate fluctuations in the combustor at a sampling frequency of 20 kHz. The PMT is equipped with a $\mathrm{CH}^*$ filter. High-speed $\mathrm{CH}^*$ chemiluminescence images are also obtained using a CMOS camera capturing 400 mm $\times$ 90 mm of the combustor onto 1200 pixels $\times$ 800 pixels of the sensor and a framing rate of 5 kHz. A total number of 7,418 images were acquired for each state of combustor operation. For this combustor, a rectangular region of size 50 mm $\times$ 50 mm after the bluff-body, as shown by the box in figure \ref{fig:Fig3_experi}(b), is used for synchronization analysis. All three measurements were obtained concurrently. More details on the dump combustor are provided in \cite{george2018pattern}.

\subsubsection{Swirl-stabilized configuration}
\begin{figure}
\centering
\includegraphics[width=120 mm]{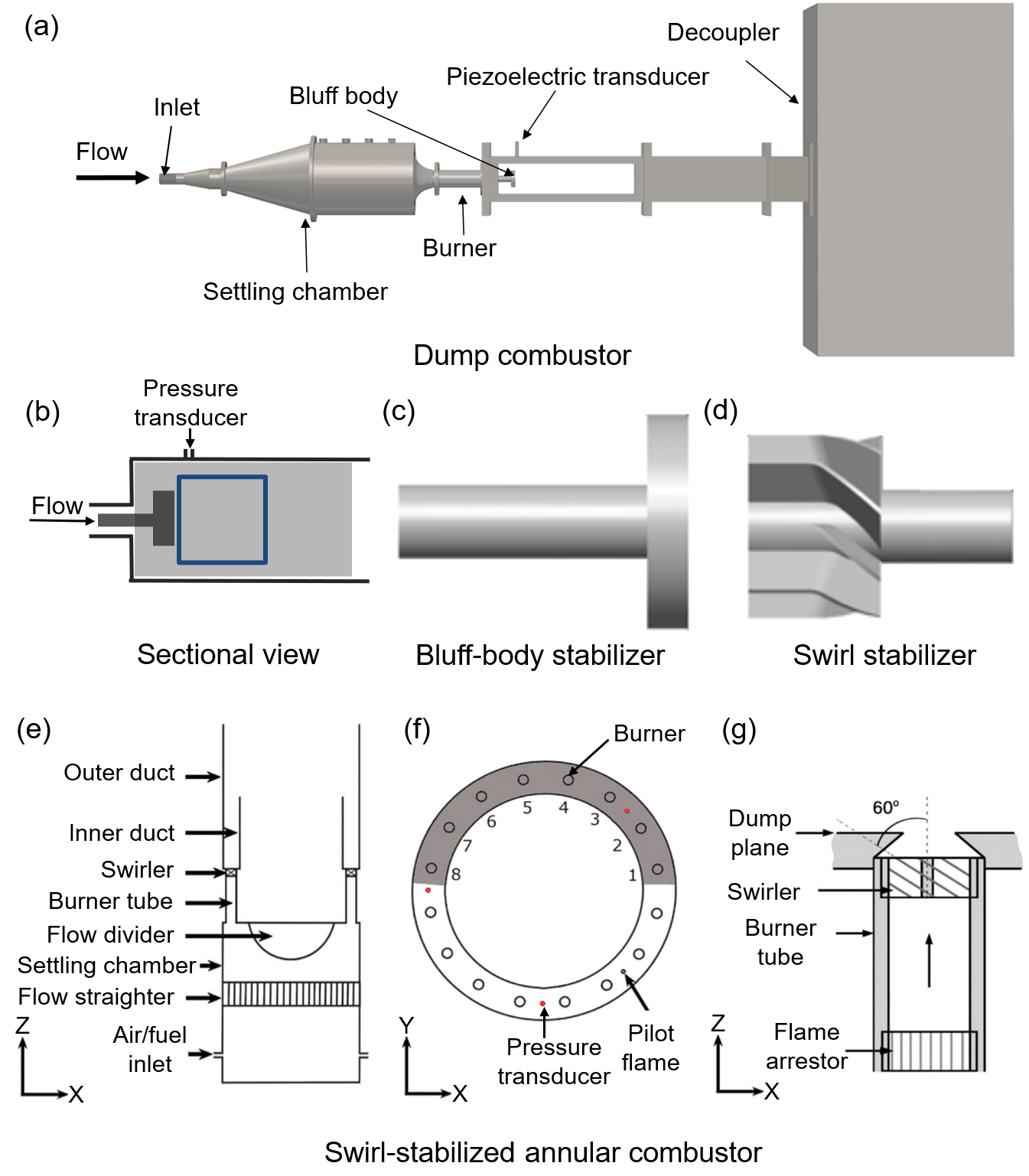}
\caption{(a) Schematic of the dump combustor. (b) Schematic of the combustor cross-section indicating the chemiluminescence field of view. We use two different flame-holding mechanisms, (c) a bluff-body and (d) swirl, attached to the burner by a central shaft. Schematic of the (e) combustor, (f) the dump plane with the burners serialized, and (g) the burner with the swirler followed by a converging section corresponding to the annular combustor. The shaded areas in (b) and (f) are captured using $\textrm{CH}^\ast$ imaging. The box in (b) is the region chosen for synchronization analysis.}
\label{fig:Fig3_experi}
\end{figure}

The swirler consists of eight guided vanes of 1 mm thickness and is mounted on a central shaft and positioned at the exit of the burner (figure \ref{fig:Fig3_experi}d). The guided vanes are inclined at $40^{\circ}$ with respect to the injector axis. The swirler has a length of 30 mm and a diameter of 40 mm. At the outer end of the swirler, a center body of diameter 16 mm and length 30 mm is attached to aid flame stabilization. In these experiments, $\dot{m}_f$ is maintained at  $7 \times10^{-4}$ kgs$^{-1}$ and $\dot{m}_a$ is varied from $0.8\times10^{-2}$ to $1.4\times10^{-2}$ kgs$^{-1}$. As a result, $\phi$ varies in the range of 0.81 to 0.53, and $Re$ varies in the range of $1.7\times10^4$ to $2.7\times10^4$. In this arrangement, only the pressure and the heat release rate fluctuations are measured using a pressure transducer and PMT at 10 kHz, respectively. 

\begin{table}
  \begin{center}
\def~{\hphantom{0}}
\begin{tabular}{lccc}
\textbf{Feature} & \textbf{Dump combustor} & \textbf{Dump combustor} & {\textbf{Annular combustor}} \\[3pt]
\hline
Flame holding mechanism & Bluff-body & Swirler & Swirler\\[3pt]
Type of bifurcation & Continuous & Secondary, abrupt & Secondary, abrupt\\[3pt]
Reynolds number, $Re$ & $2.2\times10^4$-$3.2\times10^4$ & $1.7\times10^4$-$2.7\times10^4$ & $8.6\times10^3$ \\[3pt]
Equivalence ratio, $\phi$ & 0.56 - 0.86 & 0.53 - 0.81 & 0.44 - 0.53\\[3pt]
\hline
Pressure measurement & 20 kHz & 10 kHz & 10 kHz\\
Photomultiplier tube & 20 kHz & 10 kHz & NA \\
High speed imaging & 5 kHz & NA & 2 kHz\\
\end{tabular}
\caption{\label{tab:Expt Features} Relevant properties of the turbulent combustors considered in this study.}
\end{center}
\end{table}

\subsection{Swirl-stabilized annular combustor}
The annular combustor comprising sixteen swirl-stabilized burners is shown in figure \ref{fig:Fig3_experi}(e). Technically premixed air-fuel enters the settling chamber through twelve equispaced inlet ports. The settling chamber contains honeycomb mesh to remove flow non-uniformities. A hemispherical divider guides the flow uniformly into the burner tubes. Sixteen burner tubes (length 150 mm and diameter 20 mm) connect to the annular chamber.

The arrangement of the burner tubes, location of the pilot flame and pressure measurement ports on the combustor backplane are shown in figure \ref{fig:Fig3_experi}(f). The burner tubes contain flame arresters at the bottom to prevent flashbacks and swirlers at the top. The swirling flow enters the annular combustor through a converging section. The converging nozzle has an exit diameter of 15 mm and a height of 18 mm with a contraction area ratio of 2 (figure \ref{fig:Fig3_experi}g). Sixteen swirlers are mounted on burners. The swirlers have a central shaft of 15 mm diameter on which six guide vanes are mounted at an angle of $60^{\circ}$ relative to the shaft axis. The combustion chamber is made up of two concentric ducts. The inner duct has a diameter 300 mm and a length of 200 mm, while the outer duct has a diameter and length of 400 mm. A pilot flame is used to ignite the main combustor, which is extinguished following flame stabilization in the combustor. The pilot flame is located between two injectors as shown in figure \ref{fig:Fig3_experi}f.

In these experiments, the equivalence ratio is controlled by keeping $\dot{m}_a$ at $2.86 \times 10^{-2}$ kgs$^{-1}$ fixed and varying $\dot{m}_f$ from $1.2 \times 10^{-3}$ to $1.4 \times 10^{-3}$ kgs$^{-1}$. Thus, $\phi$ is varied in the range of 0.44 to 0.53 and $Re$ is around $8.6 \times 10^{3}$. The parameter range is such that only longitudinal instability is excited, thus, allowing us to capture its dynamics with a simple one-dimensional thermoacoustic model, as done in $\S$\ref{Sec3_model}. 

A high-speed camera and piezoelectric transducer was used concomitantly to capture acoustic pressure and intensity fluctuations. The pressure signal was acquired for a duration of 3 s with a sampling rate of 10 kHz, and the camera was operated at a frame rate of 2 kHz and a resolution of 1280 $\times$ 800 pixels. A total number of 5,563 images were acquired for each state of combustor operation. Imaging is done only for one-half of the combustor backplane (see highlighted region in figure \ref{fig:Fig3_experi}f) using an air-cooled mirror located 1 m overhead the combustor.

\subsection{Equipments and experimental uncertainties}
Air ($\dot{m}_a$) and fuel ($\dot{m}_f$) flow rates are controlled using Alicat Scientific (MCR series) mass flow controller. The controller readings have an uncertainty of $\pm0.8\%$ and an additional $0.2 \%$ uncertainty of the full-scale. For all the three turbulent combustors, the uncertainty in the values of $\phi$ and $Re$ do not exceed $\pm{1.6} \%$ and $0.8 \%$, respectively. 

Piezoelectric pressure transducers (PCB103B02 Piezotronics) were used for measuring acoustic pressure fluctuations ($p^\prime$) in our experiments. The sensitivity of these transducers is 217.5 mV/kPa, along with an uncertainty of $\pm$ 0.15 Pa. The pressure fluctuations vary within $10 \%$  with respect to the reference pressure transducer for the given range of frequency over which the transducers operate. Hamamatsu H10722-01 photomultiplier tube (PMT) outfitted with $\textrm{CH}^\ast$ filter was used to measure global heat release rate fluctuations ($\dot{q}^\prime$) in the experiments. A high-speed CMOS camera (Phantom v12.1) equipped with $\textrm{CH}^\ast$ filter was used to obtain chemiluminescence images. Nikon AF Nikkor $70 - 210$ mm $f/4$ to $f/5.6$ camera lens was used to focus on the region of interest in the experiments. The $\textrm{CH}^\ast$ filter used in our experiments has a bandwidth of $435 \pm 10$ nm.

An analog-to-digital card (NI-6143, 16 bit) is used to acquire signals from the pressure transducer and PMT. To obtain concurrent measurements, Tektronix AFG1022 function generator is used for generating a pulse which then triggers the camera, and the peripheral component interconnect (PCI) card.

\section{Mean-field model of thermoacoustic transitions}
\label{Sec3_model}
\subsection{Governing equation for the acoustic field}

We begin by considering a one-dimensional thermoacoustic system assuming negligible mean flow and temperature gradient effects. In such a case, the linearized equations of momentum and energy with a heat source can be expressed as \citep{nicoud2009zero,balasubramanian2008thermoacoustic}:
\begin{subequations} 	\label{Eq-3.1}
\begin{align}
\frac{1}{\rho_0} \frac{\partial \tilde{p}^\prime}{\partial \tilde{z}} +  \frac{\partial \tilde{u}^\prime}{\partial \tilde{t}} &= 0, \label{Eq-3.1a} \\ 	\frac{\partial \tilde{p}^\prime}{\partial \tilde{t}} + \gamma p_0 \frac{\partial \tilde{u}^\prime}{\partial \tilde{z}} &= (\gamma - 1) \dot{\tilde{q}}^\prime\delta(\tilde{z}-\tilde{z}_f), \label{Eq-3.1b}
\end{align}
\end{subequations}
where $\tilde{u}^\prime$ and $\tilde{p}^\prime$ are the velocity and acoustic pressure fluctuations, respectively. Here, $\gamma$ is the ratio of specific heat capacities, $\tilde{t}$ is the time, $\tilde{z}$ is the distance along the axial direction in the duct, and $\rho_0$ and $p_0$ are the density and pressure at mean flow condition. We assume that the flame is acoustically compact and concentrated at $\tilde{z}_f$, which is indicated in (\ref{Eq-3.1b}) with the Dirac delta function $\delta(\tilde{z}-\tilde{z}_f)$.
	
We use Galerkin modal expansion to simplify the system of partial differential equations \citep{lores1973nonlinear}. We project \eqref{Eq-3.1a} and \eqref{Eq-3.1b} onto the Galerkin modes and reduce the partial differential equations to a set of ordinary differential equations. The spatially and temporally varying acoustic pressure and velocity signals are decomposed in terms of spatial basis functions (sine, cosine) satisfying appropriate boundary conditions along with time-varying coefficients ($\eta, \dot{\eta}$). Here, we choose the basis as the eigenmodes of the self-adjoint part of the linearized system. As a result, the acoustic pressure $\tilde{p}^\prime$ and the velocity fluctuations $\tilde{u}^\prime$ can be expanded as a series of orthogonal basis functions, which satisfy the boundary conditions associated with the close-open duct \citep{culick2006unsteady,nair2015reduced}:
\begin{equation}
\tilde{p}^\prime (\tilde{z},\tilde{t}) = p_0 \sum_{j = 1}^{n} \frac{\dot{\eta}_j (\tilde{t})}{\tilde{\Omega}_j} \cos(\tilde{k}_j \tilde{z}),\quad \quad \tilde{u}^\prime (\tilde{z},\tilde{t}) = \frac{p_0}{\rho_0 c_0} \sum_{j = 1}^{n} \eta_j (\tilde{t}) \sin(\tilde{k}_j \tilde{z}).
\label{Eq-3.2}
\end{equation}
Here, the time-varying coefficients $\eta_j$ and $\dot{\eta}_j$ are associated with $j^\textrm{th}$ mode of $\tilde{u}^\prime$ and $\tilde{p}^\prime$. The wavenumber and the natural frequency of the system are given by: $\tilde{k}_j = (2j-1)\pi/2\tilde{L}$ and $ \tilde{\Omega}_j = c_0 \tilde{k}_j$, respectively. By substituting \eqref{Eq-3.2} in \eqref{Eq-3.1b}, we get:
\begin{equation}
\sum_{j = 1}^{n} \frac{\ddot{\eta}_j (\tilde{t})}{\tilde{\Omega}_j} \cos(\tilde{k}_j \tilde{z})  + \frac{\gamma p_0}{\rho_0 c_0} \sum_{j = 1}^{n} \eta_j (\tilde{t}) \tilde{k}_j \cos(\tilde{k}_j \tilde{z}) = \frac{(\gamma - 1)}{p_0} \dot{\tilde{q}}^\prime\delta (\tilde{z}-\tilde{z}_f).
\label{Eq-3.3}
\end{equation}
We then project the resultant equation along the basis function by multiplying \eqref{Eq-3.3} with $\cos(\tilde{k}_j \tilde{z})$ and evaluating the inner product over the domain. Thus, we obtain a set of second-order ordinary differential equations:
\begin{equation}
\frac{\ddot{\eta}_j(\tilde{t})}{\tilde{\Omega}_j} + c_0 \tilde{k}_j \eta_j (\tilde{t}) = \frac{2(\gamma - 1)}{\tilde{L} p_0} \int_0^{\tilde{L}} \dot{\tilde{q}}^\prime \delta (\tilde{z}-\tilde{z}_f) \cos(\tilde{k}_j \tilde{z}) d\tilde{z},
\notag
\end{equation}
where, $c_0 = \sqrt{\gamma p_0 / \rho_0}$ is the average speed of sound in the duct and $\int_0^{\tilde{L}} \cos^2(\tilde{k}_j\tilde{z}) d\tilde{z} = \tilde{L}/2$.

As a first approximation, we assume that the influence of higher modes can be neglected, and a single-mode analysis captures the thermoacoustic transition reasonably well \citep{lieuwen2003modeling, culick2006unsteady, subramanian2013subcritical}. Thus, upon considering only a single mode for our analysis, we obtain:
\begin{equation}
\ddot{\eta}(\tilde{t}) + \tilde{\zeta} \dot{\eta} (\tilde{t}) + \tilde{\Omega}_0^2  \eta (\tilde{t}) = \frac{2(\gamma - 1)}{\tilde{L} p_0} \tilde{\Omega}_0 \int_0^L \dot{\tilde{q}}^\prime \delta (\tilde{z}-\tilde{z}_f) \cos(\tilde{k} \tilde{z}) d\tilde{z},
\label{Eq-3.4}
\end{equation}
where following \cite{matveev2003model}, the term $\tilde{\zeta} \dot{\eta}$ is introduced to account for acoustic damping, which plays a crucial role in determining the amplitude of limit cycle oscillation, $\tilde{\zeta}$ being the damping coefficient.

\subsection{Mean-field modeling of the heat source}
To complete the thermoacoustic model, we need to know the exact form of $\dot{\tilde{q}}^\prime$. Here, we assume that the nonlinear flame response can be approximated by a population of phase oscillators \citep{dutta2019investigating}, which evolve under the influence of acoustic fluctuations. This can be expressed in terms of a general response function ($\mathcal{G}$) which expresses the evolution of the phase of the population of oscillators as \citep{strogatz2000kuramoto,kuramoto2003chemical}:
\begin{equation}
\frac{d\theta_i(\tilde{t})}{d\tilde{t}} = \tilde{\omega}_i + \mathcal{G}\left[\hat{R}(\tilde{t}),\Phi(\tilde{t}),\theta_i(\tilde{t})\right],
\label{Eq-3.5}
\end{equation} 
where $\tilde{\omega}_i$ is the mean subtracted frequency of the $i^\textrm{th}$ oscillator, where $i=1,...,N$, and $\mathcal{G}$ is a function of the phase of the oscillators $(\theta_i)$, normalized amplitude $(\hat{R})$ and phase $(\Phi)$ of the acoustic variable. The frequencies of the oscillators are distributed according to the probability density $g(\tilde{\omega})$ centered around the acoustic frequency $\tilde{\Omega}_0$. 

The relation between $\Phi$ and $\theta_i$ can be obtained by assuming that $\mathcal{G}$ shifts $\theta_i$ closer or away from $\Phi$: $\mathcal{G}$ is positive when $\theta_i$ lags $\Phi$, and $\mathcal{G}$ is negative when $\theta_i$ leads $\Phi$. It is evident that the simplest periodic function that satisfies these requirements leads to: $\mathcal{G}\propto \sin(\Phi - \theta_i)$ \citep{kuramoto1975self}. We further assume that $\mathcal{G} \propto \hat{R}$ implies that the influence of the pressure oscillations becomes stronger as the amplitude of the pressure oscillations on the phase oscillators increases. This amplitude dependence of $\mathcal{G}$ determines the maximum rate of phase shift for a given amplitude of pressure oscillations and so determines the \textit{sensitivity} of an oscillator to the amplitude of acoustic perturbation. Thus, the modified expression for the evolution of the phase oscillator is:
\begin{equation}
\frac{d\theta_i(\tilde{t})}{d\tilde{t}} = \tilde{\omega}_i + \tilde{K} \hat{R}(\tilde{t}) \sin\left[\Phi(\tilde{t})- \theta_i(\tilde{t})\right],
\label{Eq-3.6}
\end{equation}
where the interaction amongst the oscillators is weighted equally using the coupling strength $\tilde{K}$. In the absence of acoustic feedback, $\tilde{K}$ determines the level of interaction among the oscillators and hence, controls the degree of coherence and synchrony among the oscillators. Thus, the term $\tilde{K}\hat{R}$ makes up the effective coupling strength due to intra-oscillator coupling and acoustic feedback. The distribution of frequency of the oscillators $g(\tilde{\omega})$ is estimated from the amplitude spectrum of heat release rate fluctuations during combustion noise. Hence, nonlinear flame response due to turbulence, acoustic feedback, etc., are indirectly accounted through the initial distribution $g(\tilde{\omega})$ and the mean-field interactions of the oscillators.

Finally, the individual contribution from each of the phase oscillators can be added to obtain the overall heat release rate fluctuations \citep{dutta2019investigating}, as shown below:
\begin{equation}
\dot{\tilde{q}}^\prime = \tilde{q}_0 \sum_{i=1}^N \sin \left[\tilde{\Omega}_0 \tilde{t} + \theta_i (\tilde{t})\right],
\label{Eq-3.7}
\end{equation}
where $\tilde{q}_0$ (N/ms) is introduced to keep the dimensions consistent and does not contribute to heat release rate fluctuations. The overall heat release rate fluctuations in \eqref{Eq-3.7} is thus high when oscillators are in synchrony and low otherwise, as has been corroborated in multiple studies \citep{mondal2017onset, roy2021flame}. 

\subsection{Flame-acoustic coupling}
Now we combine the mean-field model of the heat source with the oscillator equation for the temporal dynamics of the acoustic mode. Substituting (\ref{Eq-3.7}) into (\ref{Eq-3.4}), we obtain:
\begin{equation}
\ddot{\eta}(\tilde{t}) + \tilde{\zeta} \dot{\eta} (\tilde{t}) + \tilde{\Omega}_0^2  \eta (\tilde{t}) = \tilde{\beta} \tilde{\Omega}_0  \cos(\tilde{k} \tilde{z}_f) \sum_{i=1}^N \sin \left[\tilde{\Omega}_0 \tilde{t} + \theta_i (\tilde{t})\right],
\label{Eq-3.8}
\end{equation}
where $\tilde{\beta} = 2 (\gamma -1) \tilde{q}_0/\tilde{L} p_0$. We non-dimensionalize \eqref{Eq-3.8} and \eqref{Eq-3.6} using the following transformations:
\begin{equation} \label{Eq-3.9}
z = \frac{\tilde{z}}{\tilde{L}}; \quad t = \tilde{\Omega}_0 \tilde{t} ; \quad  k = \tilde{k} \tilde{L} ; \quad \zeta = \frac{\tilde{\zeta}}{\tilde{\Omega}_0};  \quad \omega_i = \frac{\tilde{\omega}_i}{\tilde{\Omega}_0}; \quad \beta = \frac{\tilde{\beta}}{\tilde{\Omega}_0}; \quad K = \frac{\tilde{K}}{\tilde{\Omega}_0}. 
\end{equation}
Using the above non-dimensionalization in \eqref{Eq-3.6} and \eqref{Eq-3.8}, we obtain:
\begin{subequations} 	\label{Eq-3.10}
\begin{align}
\ddot{\eta}(t) + \zeta \dot{\eta}(t) + \eta (t) &= \beta \cos(k z_f) \sum_{i=1}^N  \sin \left[t + \theta_i (t)\right],\label{Eq-3.10a} \\ 	
\dot{\theta}_i(t)  &= \omega_i + K \hat{R}(t) \sin\left[\Phi(t)- \theta_i(t)\right].	\label{Eq-3.10b}
\end{align}
\end{subequations}
Thus, \eqref{Eq-3.10a} and \eqref{Eq-3.10b} denote the set of $(N+1)$ coupled ordinary differential equations making up the thermoacoustic mean-field model. It is easy to observe that an increase in the effective coupling strength through an increase in $K$ or amplitude of acoustics $\hat{R}$ would lead to phase synchronization of the phase oscillators to a common phase. The synchronized state would, in turn, produce the largest driving to the damped harmonic oscillator equation governing the acoustic fluctuations through the summation in \eqref{Eq-3.10a}, consequently leading to the state of thermoacoustic limit cycle oscillations, as illustrated in figure \ref{fig:Fig1_chimera}. 

\subsection{Slow flow amplitude and phase representation of the mean-field model}
\label{subsection:slowflow}
We now seek the limit cycle solution for the set of equations given by \eqref{Eq-3.10} which will then be used for obtaining the feedback of acoustic fluctuations ($\hat{R}$, $\Phi$) on the phase oscillators. We begin by assuming that the acoustic fluctuations are quasi-harmonic such that $\eta(t)$ can be decomposed as \citep{krylov1947introduction}:
\begin{equation}
\eta (t) = - R(t)\cos\left[t + \Phi (t)\right],
\label{Eq-3.11}
\end{equation}
where $R(t)$ is the envelope amplitude and $\Phi(t)$ is the phase of the ensuing limit cycle and varies much slower than the fast time scale $2\pi/\Omega_0$. Thus, the autonomous equation describing the evolution of pressure fluctuations in \eqref{Eq-3.10a} can be decomposed further in terms of slow variables ($R$ and $\Phi$). We next calculate $\dot{\eta}(t)$ and $\ddot{\eta}(t)$ by differentiating \eqref{Eq-3.11} with respect to $t$ and substituting them in \eqref{Eq-3.10a}. After applying the method of averaging to omit the fast time scales (see $\S1$ of the Supplementary material for more details), we obtain:
\begin{equation}
\dot{R}(t) e^{i \Phi(t)} + i R(t) \dot{\Phi}(t) e^{i\Phi(t)} + \zeta R(t) e^{i \Phi (t)} =  \beta \cos(kz_f) \sum_{i=1}^N e^{i\theta_i(t)}.
\label{Eq-3.12}
\end{equation}

We multiply \eqref{Eq-3.12} by $e^{-i \Phi(t)}$ and then equate the real and imaginary parts of the equation. This gives us the evolution equations for the slowly varying amplitude and phase variables as:
\begin{subequations} 	\label{Eq-3.13}
\begin{align}
\dot{R}(t) &= \beta \cos(kz_f) \sum_{i=1}^N \cos\left[\theta_i(t) - \Phi(t)\right] - \zeta R(t), \label{Eq-3.13a} \\ 	
\dot{\Phi}(t) & = \frac{\beta}{R(t)} \cos(kz_f) \sum_{i=1}^N \sin\left[\theta_i(t) - \Phi(t)\right]. \label{Eq-3.13b}
\end{align}
\end{subequations}
These equations are the $\it{truncated}$ $\it{equations}$ \citep{balanov2008synchronization} for the evolution of $R(t)$ and $\Phi(t)$. Similarly, applying the method of averaging on \eqref{Eq-3.10b}, we observe that the equation is associated with only the slow flow variables i.e., $R$, $\Phi$, and $\theta_i$, and hence, remains unchanged. 
	
\subsection{Limit cycle solution}
\label{subsection:lco}

From \eqref{Eq-3.13a}, it is straightforward that the limit cycle solution is given by $\dot{R}(t)=0$, i.e., when the rate of change of envelope of acoustic pressure tends to zero. Although \eqref{Eq-3.13a} and \eqref{Eq-3.13b} appear compact, their analysis is quite involved as it is difficult to separate the amplitude and phase variables. To make further progress, we recast them from polar to Descartes coordinates through the following variable transformation: $A(t) = R(t) \cos\Phi(t)$ and $B(t) = R(t) \sin\Phi(t)$. Taking the time derivatives of this variable transformation and then substituting $\dot{R}$ and $\dot{\Phi}(t)$ using \eqref{Eq-3.13a} and \eqref{Eq-3.13b} followed by simplification (see Supplementary material $\S2$ for derivation), we obtain:
\begin{subequations} 	\label{Eq-3.14}
\begin{align}
\dot{A}(t)& = \beta \cos(k z_f) \sum_{i=1}^N \cos\theta_i(t) - \zeta A(t), \label{Eq-3.14a} \\ 	
\dot{B}(t) &=   \beta \cos(k z_f) \sum_{i=1}^N \sin\theta_i(t) - \zeta B(t). \label{Eq-3.14b}
\end{align}
\end{subequations}
	
Imposing the condition for limit cycle oscillations, viz., $\dot{A}(t) = 0$, $\dot{B}(t) =0$ in the equation above, the resultant amplitude of limit cycle oscillations is obtained as:
\begin{equation} \label{Eq-3.15}
 R_{\text{LCO}}(t)  = \beta \cos(k z_f) N r(t)/ \zeta,
\end{equation}
where $r(t)$, the order parameter, is defined as \citep{strogatz2000kuramoto}:
\begin{equation}
r(t) = \frac{1}{N}\sqrt{\left( \sum_{i=1}^N \sin\theta_i(t) \right)^2 + \left( \sum_{i=1}^N \cos\theta_i(t) \right)^2}.
\label{Eq-3.16}
\end{equation}
We know that during limit cycle oscillations, all the oscillators are perfectly synchronized, resulting in $r(t) \approx 1$. Thus, the amplitude of the limit cycle oscillations is expressed as $R_\textrm{LCO} = \beta \cos(k z_f)N / \zeta$. This is used for normalizing the amplitude of oscillations as $\hat{R}=R/R_\textrm{LCO}$ 

To compare with experimental observations, we normalize the model \eqref{Eq-3.10} with the amplitude of limit cycle oscillation ($R_\textrm{LCO}$). Consequently, we obtain:
\begin{subequations} 	\label{Eq-3.17}
\begin{align}
&\ddot{\hat{\eta}}(t) + \zeta \dot{\hat{\eta}}(t) + \hat{\eta} (t) = \zeta \frac{1}{N}\sum_{i=1}^N  \sin [t + \theta_i (t)],\label{Eq-3.17a} \\ 	
&\dot{\theta}_i(t)  = \omega_i + K \left\{ \dot{\hat{\eta}}(t) \cos\left[t + \theta_i(t)\right]+ \hat{\eta}(t)\sin\left[t + \theta_i(t) \right] \right\},	\label{Eq-3.17b}
\end{align}
\end{subequations}
where $\hat{\eta}(t) = \eta(t)/ R_{\textrm{LCO}}$. Similarly, $R$ and $\Phi$ in \eqref{Eq-3.10b} are rewritten in terms of $\eta$ and $\dot{\eta}$ using \eqref{Eq-3.11}.

\section{Numerical simulation}
\label{Sec4_numer_simu}
In this section, we give details about the numerical setup and parameter initialization for solving the model \eqref{Eq-3.17}. We also discuss the procedure for estimating the model parameters from experimental data.

\subsection{Initialization and bifurcation analysis}
\begin{figure}
\centering
\includegraphics[width=\linewidth]{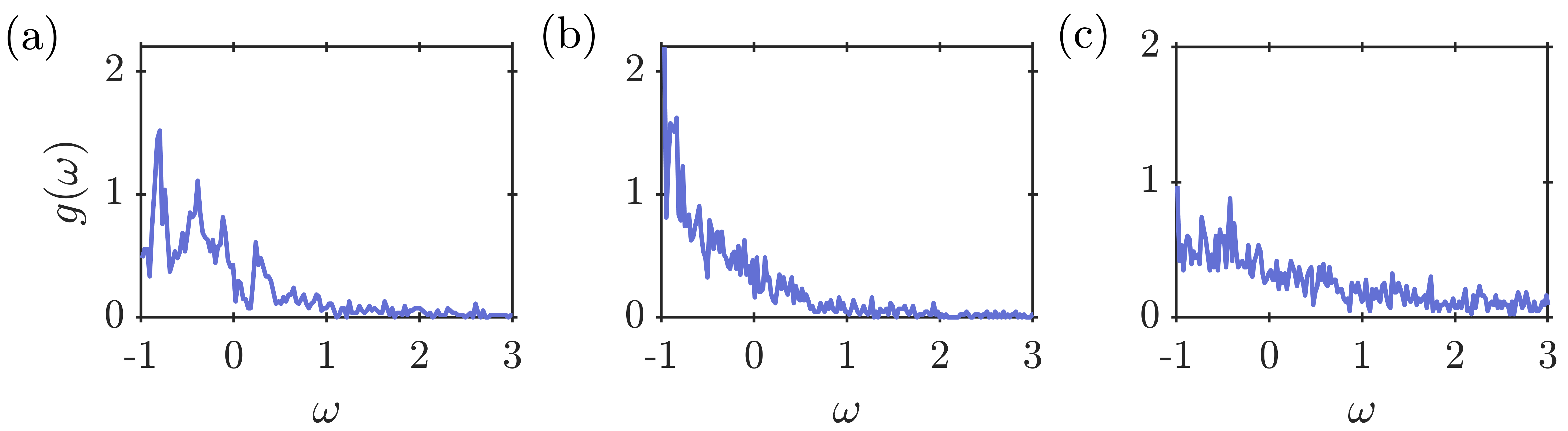}
\caption{Frequency distribution $g(\omega)$  of $N=2\times10^3$ phase oscillators obtained from the amplitude spectrum of heat release rate during the state of combustion noise for the (a) bluff-body stabilized dump combustor, (b) swirl-stabilized dump combustor and (c) annular combustor. Here, $\omega$ is distributed relative to the acoustic frequency $\Omega_0$ and also normalized by $\Omega_0$.}
\label{fig:Fig4_freq_var}
\end{figure}

The initial phase distribution is determined as $\theta_i(0) = \theta_m + \epsilon$, where $\theta_m$ represents the mean of the distribution of $\theta$ and $\epsilon$ is normally distributed as $\epsilon \backsim \mathcal{N}(0,\sigma^2)$. The initial frequency distribution $g(\omega)$ is obtained by uniformly distributing oscillators proportional to the amplitude of the heat release rate spectrum during the state of combustion noise. The procedure for numerically sampling the frequency is detailed in Appendix \ref{app_freq}. The number of oscillators $N$ determines how well the spectrum $\hat{\dot{q}}(f)$ is resolved by $g(\omega)$. We fix $N=2\times 10^3$ oscillators for which $g(\omega)$ resolves $\hat{\dot{q}}(f)$ sufficiently well and a change in $N$ does not affect the simulation results (not shown here). The initial frequency distribution of the oscillators so obtained from the heat release rate spectrum during combustion noise for the three combustors is shown in figure \ref{fig:Fig4_freq_var}.

With these inputs, the transition is obtained by sequentially changing $K$ and numerically solving \eqref{Eq-3.17a} and \eqref{Eq-3.17b} using the adaptive fourth-order Runge-Kutta method \citep{zheng2017modeling}. We first verify the transition predicted by the model with that observed in experiments. Upon verification, we perform a parameter estimation  to identify the correspondence between the control parameters in the model and our experiments, as discussed next.

\subsection{Parametric identification}
Many techniques have been used for identifying parameters while modeling thermoacoustic instabilities. For instance, system identification has been used for estimating parameters in models involving transfer functions or impulse response functions \citep{polifke2014black, bonciolini2017output} and coupled oscillator models \citep{lee2020input,lee2021system}. Similarly, uncertainty quantification has been used for estimating parameters in level-set flame models \citep{yu2019data, yu2021data}. Here, we implement an optimisation algorithm that identifies model parameters by minimizing the error between experimental data and model predictions. 

We optimize the model parameters ($\zeta$ and $K$) along with the initial conditions $\eta(0)$, $\dot{\eta}(0)$ and $\theta_i(0)$ to match the characteristics of the model output with the experimental data. Therefore, we estimate the parameter vector, $\mathbb{P}$ = $[\zeta$, $K$, $\eta(0)$, $\dot{\eta}(0)$, $\theta_m$, $\sigma]$.

Constructing the vector $\textbf{Y}$ = $[\eta, \dot{\eta},\theta_i]^T$ and re-writing the nonlinear set of equations in \eqref{Eq-3.17} as:
\begin{equation} \label{Eq-4.1}
\dot{\textbf{Y}} = f(\mathbb{P},\textbf{Y}).
\end{equation}
Starting from an initial state $\textbf{Y}_0$ at $t_0$, the state of the system at time $t_n=t_0+n\Delta t$ can be obtained from the model as:
\begin{equation} \label{Eq-4.2}
\textbf{Y}_\textrm{m}(\mathbb{P};t_n) = \int_{t_0}^{t_n} f(\mathbb{P},\textbf{Y}) dt + \textbf{Y}_0.
\end{equation}
Fourth-order Runge-Kutta scheme is used for numerically integrating the above equations with a time step of $\Delta t$. Using \eqref{Eq-3.2} and \eqref{Eq-3.7}, we obtain:
\begin{equation} \label{Eq-4.3}
p^\prime_\textrm{m} = \dot{\hat{\eta}}(t); \quad \quad \dot{q}^\prime_\textrm{m} = \frac{1}{N} \sum_{i=1}^N \sin \left[t + \theta_i(t)\right].
\end{equation}
Thus, from the model we obtain: $\textbf{X}_\textrm{m}(\mathbb{P};t_n)$= $[p^\prime_\textrm{m},\dot{q}^\prime_\textrm{m}]$. 

Next, from the experimental data, we construct the vector $\textbf{X}_\textrm{exp}(\phi; t_n)$= $[p^\prime$, $\dot{q}^\prime]^T$, where $\phi$ is equivalence ratio or the experimental control parameter, $p^\prime$ and $\dot{q}^\prime$ are the normalized acoustic pressure and global heat release rate fluctuations, respectively. Thus, we can obtain the parameter values that minimize the error between the model $\textbf{X}_{\text{m}}$ and the experiments $\textbf{X}_{\text{exp}}$. This minimisation of the error can be done by constructing the loss function ($\mathcal{L}$) based on the mean square error: 
\begin{equation} \label{Eq-4.4}
\mathcal{L}(\mathbb{P}) = \frac{1}{N} \sum_{n=1}^N \left \lVert \textbf{X}_\textrm{m}(\mathbb{P};t_n) - \textbf{X}_\textrm{exp}(\phi;t_n) \right \rVert_2^2.
\end{equation}
  
Finally, the parameter estimation can be cast in terms of a minimisation problem subject to the parameter vector $\mathbb{P}$. The minimisation is performed using the gradient descent method \citep{boyd2004convex}:
\begin{equation}
\mathbb{P}_{i+1} = \mathbb{P}_i - \alpha_l \nabla_\mathbb{P} \mathcal{L},
\label{Eq-4.5}
\end{equation}
where $\alpha_l$ is the learning rate which gives the rate at which the parameter updates per unit gradient of the loss function with respect to the parameter. We used a learning rate of $\alpha_l= 1 \times 10^{-3}$ to optimize the parameter estimates. We use the automatic differentiation method to evaluate the gradient of $\mathcal{L}$ with respect to $\mathbb{P}$ \citep{baydin2018automatic}. Kindly see Appendix \ref{app_loss} for more details about convergence and loss minimisation.

\section{Model prediction of transition to thermoacoustic instability}
\label{Sec5_model_pred}

The model is numerically implemented by choosing the frequency distribution of oscillators $g(\omega)$ from the amplitude spectrum $\hat{\dot{q}}(f)$ of the heat release rate fluctuations obtained during the state of combustion noise from different combustors. The damping coefficient ($\zeta$) is obtained from the experimental data during the state of combustion noise using parameter optimisation \eqref{Eq-4.5} and is subsequently fixed for determining other states during the transition. Using these inputs, the transition to the state of limit cycle is obtained by increasing the coupling strength $K$. Once the transition is qualitatively verified with the experiments, parameter optimisation \eqref{Eq-4.5} is performed for all the experimentally observed states to identify the relationship between the coupling strength $K$ and the control parameter $\phi$ used in experiments.

Let us now compare how our model fares in predicting continuous and abrupt transitions observed in turbulent thermoacoustic systems. 

\subsection{Continuous transition to thermoacoustic instability}
\label{S5.1-continuous}

\begin{figure}
\centering
\includegraphics[width=\textwidth]{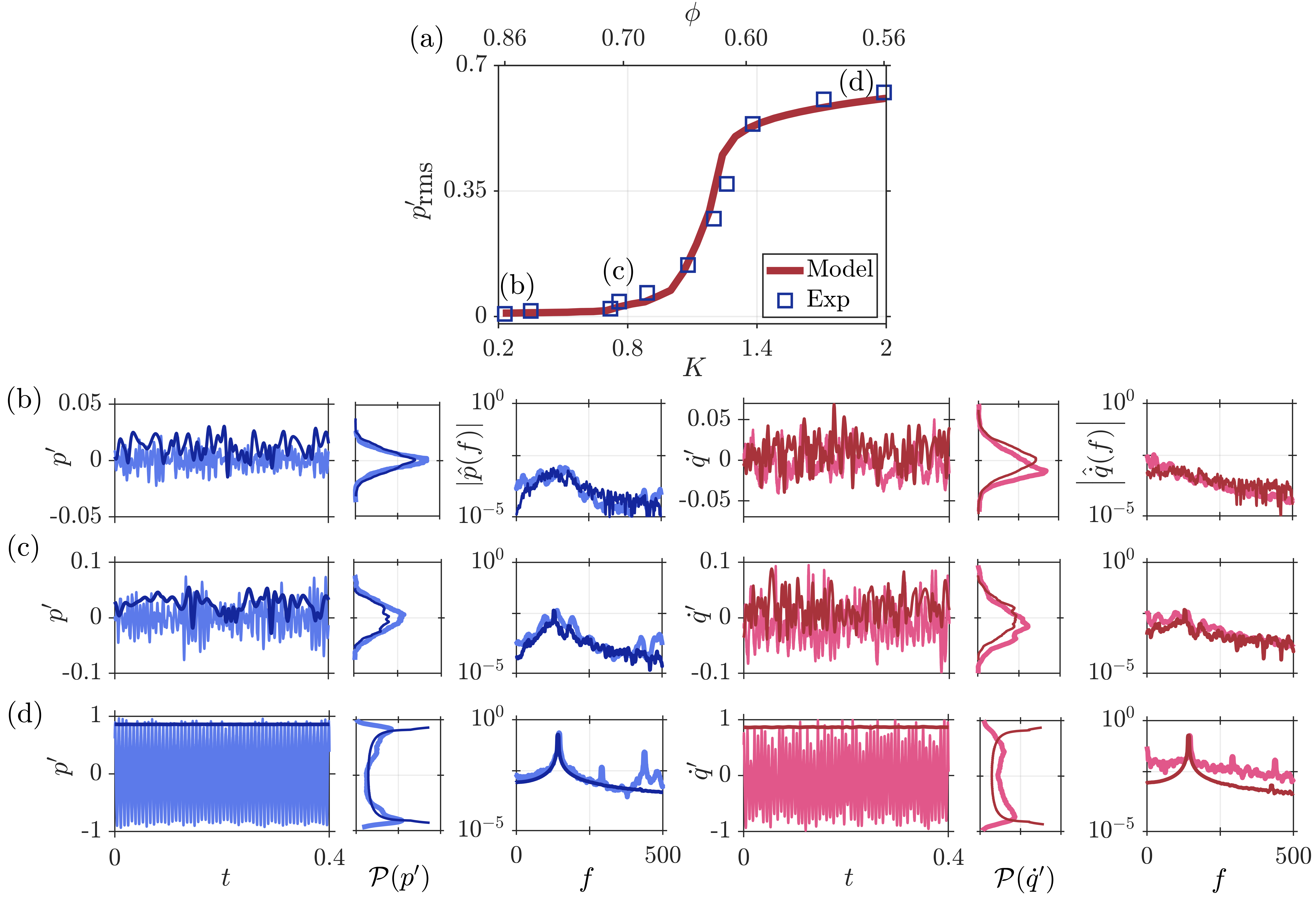}
\caption{Continuous transition to thermoacoustic instability through the state of intermittency observed in the bluff-body stabilized combustor. (a) Comparison of the normalized amplitude $p^\prime_{\text{rms}}$ obtained from the experiments ($\Box$) and model (--) as a function of the equivalence ratio ($\phi$) and the model parameter ($K$). The time series ($p^\prime$, $\dot{q}^\prime$), probability distribution function [$\mathcal{P}(p^\prime), \mathcal{P}(\dot{q}^\prime)$] and amplitude spectrum [$|\hat{p}(f)|,|\hat{\dot{q}}(f)|$] are shown during the states of (b) combustion noise ($\phi= 0.86$, $K \approx 0.23$), (c) intermittency ($\phi= 0.72$, $K \approx 0.75$) and (d) thermoacoustic instability ($\phi= 0.56$, $K \approx 2$). In panels (b-d) the experimental data are shown using a lighter shade, while the model result is shown using a darker shade. Only the envelope of the time series from the model is shown for clarity.}
\label{fig:Fig5_continuous}
\end{figure}

The bluff-body stabilized combustor, shown in figure \ref{fig:Fig3_experi}(a,c), exhibits a continuous transition to the state of thermoacoustic instability through intermittency when the equivalence ratio $\phi$ is decreased. Figure \ref{fig:Fig5_continuous}(a) shows the variation in the amplitude of acoustic pressure fluctuations ($p^\prime_{\text{rms}}$) as a function of the equivalence ratio $\phi$ and coupling strength $K$. The amplitude is normalized with the amplitude of limit cycle oscillations to aid comparison of the observed transition with that predicted by the model \eqref{Eq-3.17}. Figure \ref{fig:Fig5_continuous}(b-d) shows the characteristics of pressure ($p^\prime$) and heat release rate ($\dot{q}^\prime$) fluctuations during the state of combustion noise, intermittency and thermoacoustic instability, as marked in the bifurcation diagram. 

At $\phi = 0.86$ ($K \approx 0.23$), we observe the occurrence of combustion noise (figure \ref{fig:Fig5_continuous}b) where the fluctuations ($p^\prime, \dot{q}^\prime$) are visibly aperiodic. These fluctuations are associated with a unimodal distribution and possess broadband spectra. As $\phi$ is decreased ($\phi=0.72$, $K \approx 0.75$), we observe intermittency where periodic fluctuations appear amidst aperiodic fluctuations (figure \ref{fig:Fig5_continuous}c). A further decrease in $\phi$  leads to an increase in the frequency of occurrence and the amplitude of these periodic bursts. The appearance of these periodic bursts lead to a sharp, albeit low-amplitude, peak in the spectrum $|\hat{p}(f)|$ and $|\hat{\dot{q}}(f)|$. Further decrease in the value of $\phi$ leads to a gradual increase in the amplitude and the number of occurrences of these bursts and manifests as a continuous increase in the value of $p^\prime_{\text{rms}}$ observed in figure \ref{fig:Fig5_continuous}(a). The transition has a ``sigmoid'' shape which is a characteristic feature of such continuous transitions \citep{nair2014intermittency}. Finally, at $\phi = 0.56$ ($K \approx 2$), we observe limit cycle oscillations associated with the state of thermoacoustic instability. The fluctuations possess a bimodal distribution and are periodic with narrowband amplitude spectra at a frequency of 146.5 Hz (figure \ref{fig:Fig5_continuous}d).

The transition predicted by the model (continuous line) is also shown in figure \ref{fig:Fig5_continuous}(a). To aid comparison of specific states depicted by the model, the envelopes of the time series obtained from the model (in darker shade) are overlaid on top of each of the time series obtained from experiments (in lighter shade) in figure \ref{fig:Fig5_continuous}(b-d). Similarly, the PDF and the spectra indicated in the darker shade are the predictions of the model. The broadband spectrum $|\hat{\dot{q}}(f)|$ during the occurrence of combustion noise shown in figure \ref{fig:Fig5_continuous}(b) is used for obtaining the distribution $g(\omega)$ shown in figure \ref{fig:Fig4_freq_var}(a). 

We notice that the model captures many features of the experimental data. Foremost, we observe that the model predicts the transition observed in the experiments very well. The continuous, sigmoid-type transition to the state of limit cycle oscillation is well captured (figure \ref{fig:Fig5_continuous}a). Quite notably, the prediction from this model shows a qualitative match with the time series of pressure fluctuations obtained from experiments. The model captures various features of the time series of pressure fluctuations.  For instance, the distribution $\mathcal{P}(p^\prime)$ and spectrum $|\hat{p}(f)|$ during the state of combustion noise and intermittency are well approximated. Similarly, the envelope of limit cycle obtained from the model is a good estimate of the limit cycle amplitude observed in figure \ref{fig:Fig5_continuous}(d). Finally, the spectrum $|\hat{p}(f)|$ from the model shows a close match with that observed in experiments. 

\begin{figure}
\centering
\includegraphics[width=\textwidth]{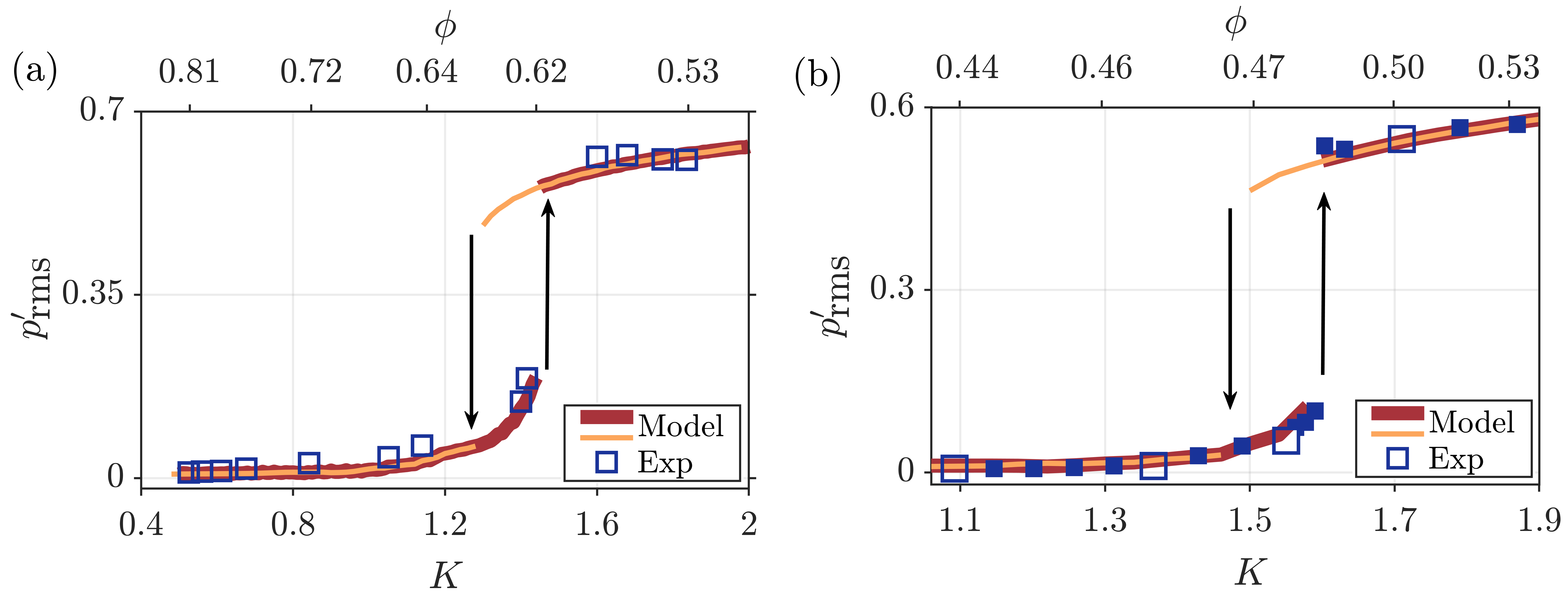}
\caption{Abrupt secondary bifurcation to limit cycle oscillations in (a) swirl-stabilized dump combustor and (b) annular combustor. The bifurcation plots illustrate the comparison of the normalized amplitude $p^\prime_{\text{rms}}$ obtained from the experiments ($\Box$) and the model (--) as a function of equivalence ratio ($\phi$) and the model parameter ($K$). The forward and reverse transitions in the model are indicated as (\textbf{--}) and (--), respectively. In panel (b), heat release rate data is available only at the four points indicated by ($\Box$). For the remaining points (depicted by $\blacksquare$) only pressure data is available, we interpolated these points from the optimisation at the four points shown in figure \ref{fig:Fig8_param_relation}.}
\label{fig:Fig6_secondary_bif}
\end{figure}
% The parameter relation ($\phi-k$) during the remaining points (depicted by $\circ$) for which only pressure data is available are interpolated from the optimisation at the four points.

\subsection{Secondary bifurcation to high-amplitude thermoacoustic instability}
\begin{figure}
\centering
\includegraphics[width=\textwidth]{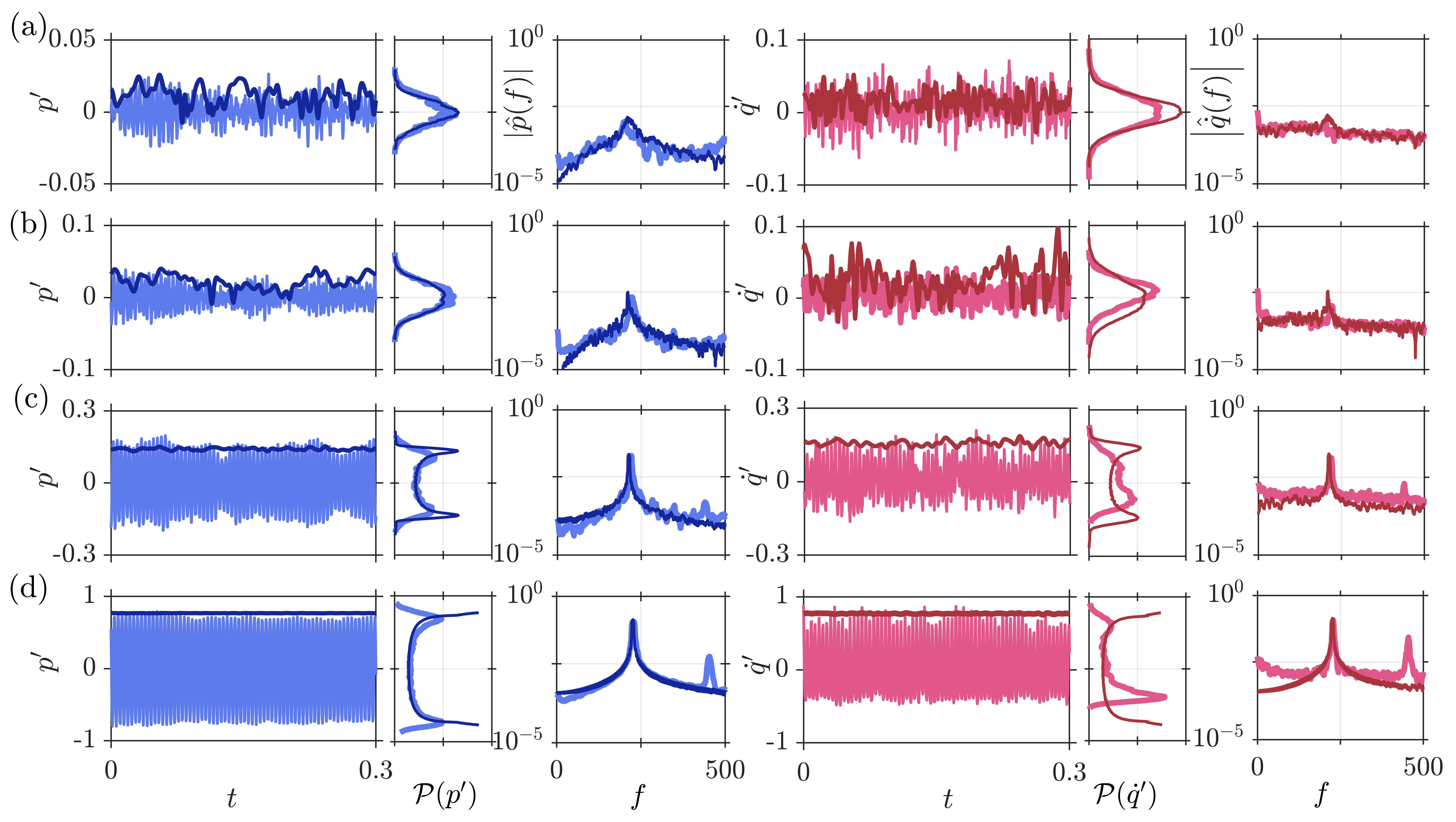}
\caption{Representative states observed during secondary bifurcation in the annular combustor. The time series ($p^\prime$, $\dot{q}^\prime$), probability distribution function [$\mathcal{P}(p^\prime), \mathcal{P}(\dot{q}^\prime)$] and amplitude spectrum [$|\hat{p}(f)|,|\hat{\dot{q}}(f)|$] are shown during the states of (a) combustion noise ($\phi=0.44, K= 1.09$), (b) intermittency ($\phi=0.47, K= 1.37$), (c) low-amplitude thermoacoustic instability ($\phi=0.49, K= 1.55$) and (d) high-amplitude thermoacoustic instability ($\phi=0.52, K= 1.71$). The experimental data are shown using a lighter shade, while the model result is shown using darker shade. For clarity, only the envelope of time series from the model is included.}
\label{fig:Fig7_seco_timeseries}
\end{figure}

Both swirl-stabilized dump combustor (figure \ref{fig:Fig3_experi}a, d) and annular combustor (figure \ref{fig:Fig3_experi}e) undergo abrupt secondary bifurcation to the state of thermoacoustic instability on varying the control parameter ($\phi$) systematically (see figure \ref{fig:Fig6_secondary_bif}). For the former, the transition is observed when $\phi$ is decreased from 0.81 to 0.53, while for the latter, the transition is observed upon increasing $\phi$ from 0.44 to 0.53. The transition for both the combustors are shown in figure \ref{fig:Fig6_secondary_bif} where the normalized root mean square of the acoustic pressure ($p^\prime_\textrm{rms}$) is plotted as a function of the equivalence ratio ($\phi$) and the model parameter $K$. For the annular combustor (figure \ref{fig:Fig6_secondary_bif}b), chemiluminescence data was acquired only at four locations marked by $\Box$. For the remaining states, pressure data alone was acquired. The dynamical states corresponding to the four points representative of the secondary bifurcation in the annular combustor are shown in figure \ref{fig:Fig7_seco_timeseries}.

%\begin{figure}
%\centering
%\includegraphics[width=\textwidth]{seco_ts.png}
%\caption{Representative states observed during secondary bifurcation in the swirl-stabilized turbulent combustor. The time series ($p^\prime$, $\dot{q}^\prime$), probability distribution function [$\mathcal{P}(p^\prime), \mathcal{P}(\dot{q}^\prime)$] and amplitude spectrum [$|\hat{p}(f)|,|\hat{\dot{q}}(f)|$] are shown during the states of (a) combustion noise ($\phi=0.81, K= 0.53$), (b) intermittency ($\phi=0.66, K= 1.05$), (c) low-amplitude thermoacoustic instability ($\phi=0.62, K= 1.41$) and (d) high-amplitude thermoacoustic instability ($\phi=0.58, K= 1.66$). The experimental data are shown using a lighter shade, while the model result is shown using darker shade. Only the envelope of the time series from the model is shown for clarity.}
%\label{fig:Fig7_seco_timeseries}
%\end{figure}

The annular combustor is in a state of combustion noise close to $\phi$ = 0.44 (figure \ref{fig:Fig7_seco_timeseries}a). Both pressure and heat release rate fluctuations exhibit noisy behavior, possessing a unimodal PDF and broadband amplitude spectrum. When $\phi$ is increased, we observe intermittent periodic oscillations (figure \ref{fig:Fig7_seco_timeseries}b). The appearance of intermittent bursts is associated with a continuous increase in the amplitude of oscillations (see figure \ref{fig:Fig6_secondary_bif} for $\phi\approx 0.47$ and $K \approx$ 1.37). In addition, the peak of the distribution $\mathcal{P}(p^\prime)$ and $\mathcal{P}(\dot{q}^\prime)$ widens. A narrowband starts to appear in the amplitude-spectrum at $f= 218$ Hz (figure \ref{fig:Fig7_seco_timeseries}b). Upon further increase to $\phi \approx$ 0.49 ($K \approx$ 1.55), we observe low-amplitude limit cycle oscillations along with a bimodal distribution $\mathcal{P}(p^\prime)$ possessing additional peaks at $|p^\prime|\neq 0$ (figure \ref{fig:Fig7_seco_timeseries}c). During the state of low-amplitude limit cycle oscillations, we observe $\tilde{p}^\prime_{\text{rms}}\approx 0.3$ kPa. The distribution $\mathcal{P}(\dot{q}^\prime)$ depicts distinct bimodal shape along with a narrowband peak in the amplitude-spectrum at $f= 223$ Hz. Finally, there is an abrupt secondary fold bifurcation from low-amplitude to very high-amplitude limit cycle oscillations beyond $\phi \approx 0.49$ ($K \approx$ 1.55 in figure \ref{fig:Fig6_secondary_bif}b). The amplitude of pressure fluctuations at this state is $\tilde{p}^\prime_\textrm{rms}\approx 2$ kPa. The distribution remains bimodal with the distribution peaks appearing at large values of $p^\prime$ (figure \ref{fig:Fig7_seco_timeseries}d). Accordingly, the spectrum shows a dominant peak at $f=227$ Hz.

A similar transition is also observed in the swirl-stabilized dump combustor (see figure \ref{fig:Fig6_secondary_bif}a) when $\phi$ is decreased from $0.81$ to $0.53$. There is a secondary bifurcation from low-amplitude ($\tilde{p}^\prime\approx 1.5$ kPa) limit cycle to very high-amplitude ($\tilde{p}^\prime\approx 5.6$ kPa). The frequency of limit cycle oscillations is $f = 201$ Hz. The dynamical states during the secondary bifurcation are similar to those observed for the annular combustor that is depicted in figure \ref{fig:Fig7_seco_timeseries}.

Secondary bifurcation predicted by the model for each of the two combustors is also shown in figure \ref{fig:Fig6_secondary_bif}. We notice that the model, numerically simulated by considering the spectrum $|\hat{\dot{q}}(f)|$ during the state of combustion noise for obtaining $g(\omega)$ (see figure \ref{fig:Fig7_seco_timeseries}a), predicts the secondary bifurcation very well. The normalized limit cycle amplitude following the secondary fold bifurcation is well approximated. The dynamics of pressure and heat release fluctuations obtained from the model are also shown in figure \ref{fig:Fig7_seco_timeseries}. We observe that the model captures the amplitude of pressure and heat release rate oscillations associated with the different states of combustor operation. In addition, the probability density functions and amplitude spectrum of the pressure and the heat release rate fluctuations are well estimated by the model.

\subsection{Relation between experimental and model parameter}

\begin{figure}
\centering
\includegraphics[width=\textwidth]{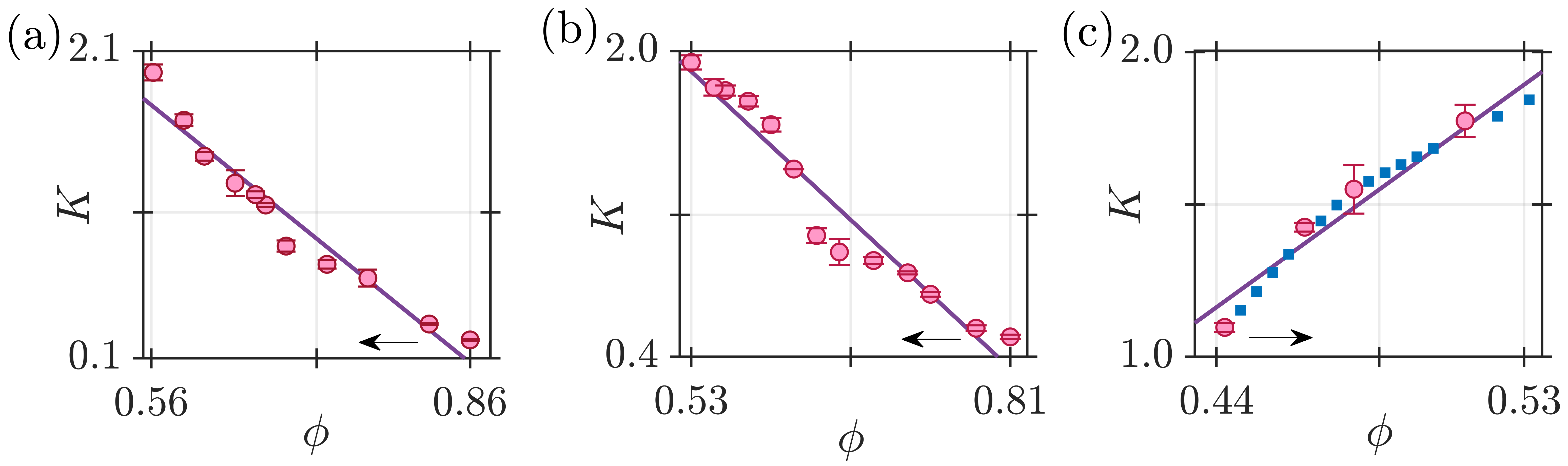}
\caption{Correspondence between the model parameter ($K$) and the experimental control parameter ($\phi$) obtained using gradient descent optimisation for the (a) bluff-body stabilized dump combustor, (b) swirl-stabilized dump combustor, and (c) annular combustor. For all the cases, we obtain a linear relation  between $\phi$ and $K$ where the fits are:  (a) $K = -5.35\phi + 4.68 $, (b) $K = -4.85\phi + 4.28 $ and (c) $K = 8.59\phi - 2.66$. The goodness-of-fit corresponding to different combustors is 0.95, 0.94, and 0.96, respectively. The arrows in each panel indicate the direction in which $\phi$ was varied in experiments. In panel (c), heat release rate data were obtained only at four points. The remaining points were interpolated. Error bars correspond to the standard deviation in the estimation of $K$ by sliding the window used during optimisation (Appendix \ref{app_freq}).}
\label{fig:Fig8_param_relation}
\end{figure}

The correspondence between the control parameters in experiments and the parameters of the model is quite important. Knowing how experimental parameters are related to the model allows for the interpretation of experimental observations in terms of the physics embodied in the model. Thus, we perform parameter optimisation as explained in \S\ref{Sec4_numer_simu} to obtain the relation between the experimentally controlled equivalence ratio $\phi$ and model parameter $K$. The relationship between $K$ and $\phi$ estimated using parameter optimisation for the three combustors are shown in figure \ref{fig:Fig8_param_relation}. The indicated error in the relation is determined from a distribution of $K$ obtained by estimating $K$ from $\textbf{X}_{\textrm{exp}}$ for a window of size $t_\textrm{w}=0.5$ s and sliding the window across the entire time series. The choice of window size is explained in Appendix \ref{app_freq}. These $K-\phi$ relations were used for constructing the bifurcation diagram in terms of $K$ in figures \ref{fig:Fig5_continuous}a and \ref{fig:Fig6_secondary_bif}. Note that for the annular combustor, optimisation was performed only at four data points for which heat release rate data was available. 

%The resulting $\phi-K$ relationship was then used for interpolating the remaining points (marked by $\circ$ in figure \ref{fig:Fig6_secondary_bif}b).

% The parameter relation ($\phi-k$) during the remaining points (depicted by $\circ$) for which only pressure data is available are interpolated from the optimisation at the four points.
%The parameter optimisation for the continuous transition observed in bluff-body combustor indicates that $\phi$ depends on $K$ through a simple linear relationship: $\phi = -5.35K + 4.68 $ (figure \ref{fig:Fig8_param_relation}a). For the secondary bifurcation (figure \ref{fig:Fig8_param_relation}b,c), the relationship for the swirl-stabilized combustor is $\phi = -4.85K + 4.28 $ and for annular combustor, it is $\phi= 8.59K-2.66$. The goodness-of-fit corresponding to different combustors is 0.95, 0.94, and 0.96, respectively. 
For the bluff-body and swirl-stabilized dump combustors, the transition is attained by decreasing $\phi$. For each of these cases, $\phi$ is a linearly decreasing function of $K$. In contrast, the transition is attained by increasing $\phi$ in the annular combustor, and we obtain a linearly increasing relationship between $\phi$ and $K$. For all these cases, note that the linear relation between $\phi$ and $K$ is such that (increasing/decreasing) change in the control parameter $\phi$ is translated to an increase in $K$. This linear dependence makes the model highly interpretable: a change in the control parameter leads to an increase in the coupling strength of the phase oscillators, promoting global phase synchronization and hence, limit cycle oscillations. 
%Thus, the model is easily interpretable in terms of experimentally relevant control parameters. 

Here, we reiterate that the excellent agreement between the experiments and model is not a result of parameter optimisation. The parameter optimisation only determines the specific mapping from $K$ to $\phi$ once the transition in the model has been determined for a given parameter input. The accuracy of prediction and optimisation depends upon how well the model represents the behavior of the system. Our results show that the model captures the combustor dynamics very well, lending support to our methodology.

On the flip-side however, the model does not capture some experimentally observed features. For instance, we observe that the higher order modes appearing in the spectrum $|\hat{p}(f)|$ (cf. figures \ref{fig:Fig5_continuous} and \ref{fig:Fig7_seco_timeseries}) are not captured, as only the fundamental mode was considered in the formulation of the model \eqref{Eq-3.4}. Further, the presence of highly turbulent flow in the experiments makes it difficult for the model to capture the exact characteristics of $\dot{q}^\prime$ fluctuations during different states of combustor operation (cf. figures \ref{fig:Fig5_continuous} and \ref{fig:Fig7_seco_timeseries}). Hence, model estimates of $\dot{q}^\prime$ can potentially be improved by introducing stochastic terms in \eqref{Eq-3.10}. %Further, the model also captures deterministic behavior, such as the observation of multifractality during combustion noise (not shown here). 
For the present discussion in the context of modeling and prediction of thermoacoustic transitions along with the underlying synchronization behavior, these dissimilarities are ignored in view of simplicity and parsimony. 

To understand the reason for the observation of continuous and abrupt transition across disparate combustion systems under seemingly identical operating conditions, we next consider the characteristics of synchronization in more detail.

\section{Continuous and explosive synchronization transition to thermoacoustic instability}
\label{Sec6_Synchor}
Let us now consider the characteristics of synchronization which underlies the two different kinds of transitions in more detail. To make this connection substantive, we compare the behavior of phase oscillators in the mean-field model devoid of any spatial inputs with the phase dynamics of spatially distributed heat release rate oscillations obtained through chemiluminescence imaging in our experiments. 

Figure \ref{fig:Fig1_chimera} shows the instantaneous spatial distribution of phasors ($\psi_i =  \theta_i - \Phi$) for the bluff-body stabilized combustor and the annular combustor, where $\theta_i$ and $\Phi$ are the instantaneous phase of $\dot{q}^\prime(x,y,t)$ and $p^\prime(t)$. The phase difference is related to the correlation over a small time window of the time series \citep{sethares2007rhythm,balasubramanian2008thermoacoustic}:
\begin{equation}
\cos\psi_i = \int_0^t p^\prime(t^\prime)\dot{q}^\prime(x,y,t^\prime)dt' \bigg/ \left[\int_0^t \left(p^\prime(t')\right)^2 dt' \int_0^t \left(\dot{q}^\prime(x,y,t')\right)^2dt'\right]^{1/2},
\label{Eq-Rayleigh-Phase}
\end{equation}
which is the local Rayleigh index and is related to the acoustic power added to the acoustic field due to the flame fluctuations. Here, we measure the phase of heat release rate fluctuations from arbitrary pixels in chemiluminescence images and pressure fluctuations using the Hilbert transform. To reduce the effect of noisy fluctuations, the chemiluminescence images obtained from the bluff-body and annular combustor are coarse-grained over $8\times 8$ pixels and $6 \times 6$ pixels, respectively. While the signals $p^\prime(t)$ and $\dot{q}^\prime(x,y,t)$ at various conditions are not always strictly analytic, the Hilbert transform can still be used for the purposes of visualization. Indeed, \cite{mondal2017onset} explicitly evaluated the correlation \eqref{Eq-Rayleigh-Phase} as well as used the probability of recurrence to determine the phase of the heat release rate field. The resulting phases were qualitatively similar to the phase obtained through the Hilbert transform. Hence, we adopt the same in the following. 

In addition, we also ensured that there were no acoustic phase delay effects in our experimental measurements. To this end, the pressure oscillations are measured at 25 mm from the dump plane in the bluff-body combustor and remain nearly constant across the domain of spatial measurements (see figure \ref{fig:Fig3_experi}b). In the case of the annular combustor, the acoustic pressure is measured on the combustor backplane such that there is no acoustic phase delay (cf. figure \ref{fig:Fig3_experi}f). We also mask out the regions between the swirling flames in the annular combustor and do not consider their contribution in our calculations. This is done because heat release rate fluctuations from the inter-flame region are not significant and remain noisy for all the dynamical states (cf. figure \ref{fig:Fig1_chimera}d-f).

Figure \ref{fig:Fig1_chimera} illustrates the distinct patterns in the phase-field obtained from experiments. During the occurrence of combustion noise, the phase-field is randomly oriented and incoherent (cf. figure \ref{fig:Fig1_chimera} a,d). This state is more generally referred to as  \textit{phase turbulence} \citep{shraiman1986order, shraiman1992spatiotemporal}. In the present context, the phase-turbulent state indicates that the heat release rate response of the flame is dominated only by the highly turbulent flow, leading to an incoherent and desynchronized field of the phase difference between the pressure and the heat release rate fluctuations. As the phasors are distributed $\psi_i>|\pi/2|$, the acoustic power production remains very low \eqref{Eq-Rayleigh-Phase}, resulting in low-amplitude aperiodic fluctuations. In contrast, during the state of thermoacoustic instability, the phasors are aligned in a coherent manner, highlighting the global phase synchronization (cf. figure \ref{fig:Fig1_chimera} c,f). The phasors distributed in $\psi_i<|\pi/2|$, effect substantially high acoustic power production driving thermoacoustic instability in the two combustors.

While there are similarities in the characteristics of phase turbulence and phase synchronization during combustion noise and thermoacoustic instability, the manner in which synchronization is attained is quite different. For the bluff-body stabilized and annular combustor, the emergence of a globally synchronized state takes place through the state of intermittency where the phase-field shows both phase turbulence and phase synchronization, as can be observed in figure \ref{fig:Fig1_chimera}(b,e). Thus, clusters of synchronized regions appear amidst phase turbulence. The states of co-existence of clusters of phase turbulence and synchronization are referred to as \textit{chimeras} \citep{kuramoto2002coexistence, abrams2004chimera}. On the other hand, for the annular combustor, the transition takes place through the state of intermittency and low-amplitude instability. The phase-field is shown for the state of intermittency in figure \ref{fig:Fig1_chimera}(e). 
%We again notice the co-existence of clusters of phase turbulence and synchronization, indicating the presence of chimera states during low-amplitude instability. 

To quantify the characteristics of synchronization, we define the Kuramoto order parameter:
\begin{equation}
\bar{r} = \bigg\langle \bigg|\frac{1}{N}\sum_{i=1}^N \exp\left(i\theta_i(t)\right)\bigg|\bigg\rangle_t, \qquad \bar{r}\in[0,1],
\label{Eq-Kuramoto-Order}
\end{equation}
where $\langle \cdot\rangle_t$ implies time average and $\theta_i$ is the phase at $i$-th oscillator. The order parameter quantifies the degree of synchrony among the oscillators as the bifurcation parameter varies. Kuramoto order parameter is close to zero for desynchronized states and is close to one for synchronized states.

\begin{figure}
\centering
\includegraphics[width=\linewidth]{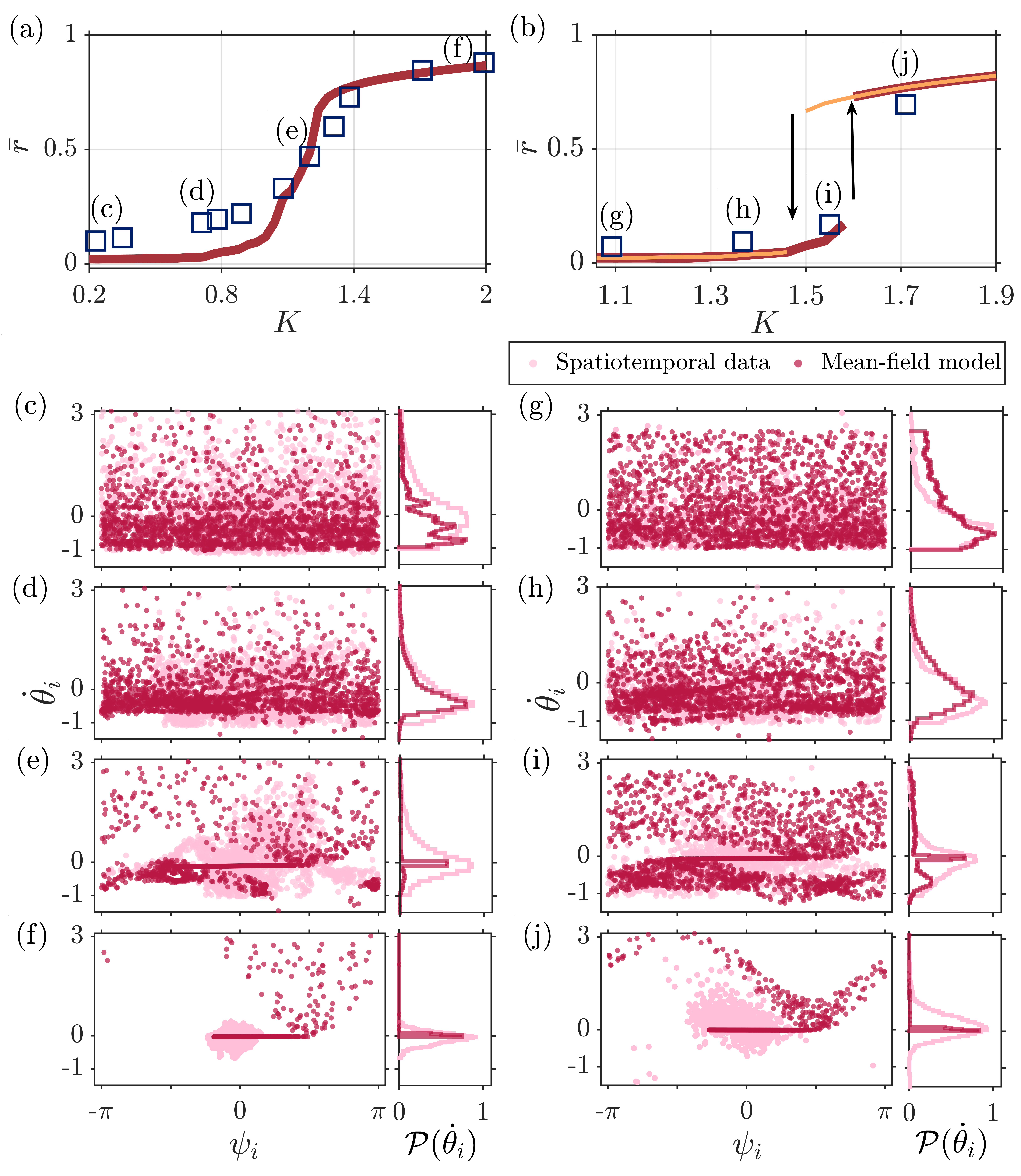}
\caption{The variation of time-averaged Kuramoto order parameter $\bar{r}$ \eqref{Eq-Kuramoto-Order} as a function of coupling strength $K$ shows (a) a continuous \textit{second-order} synchronization transition for the bluff-body stabilized combustor and (b) abrupt \textit{first-order} explosive synchronization transition for the annular combustor. Panels (c-j) depict the instantaneous oscillator distribution in the $\dot{\theta}_i-\psi_i$ phase space along with the distribution $\mathcal{P}(\dot{\theta}_i)$ representative of various dynamical states indicated in panels (a) and (b). Supplementary movies S1-S8 depict the time evolution of oscillators during various states. In panels (c-f), the distribution of oscillators obtained from spatiotemporal imaging of the combustors is depicted using a lighter shade, while the oscillators from the mean-field model using a darker shade. Here, $\dot{\theta}_i$ is normalized, and the mean acoustic frequency $\Omega_0$ has been subtracted.}
\label{fig:Fig9_order_param}
\end{figure}

Figure \ref{fig:Fig9_order_param}(a,b) illustrates the variation of the order parameter when the coupling strength $K$ is varied. The order parameter is determined according to \eqref{Eq-Kuramoto-Order} from the model \eqref{Eq-3.17}. The procedure for obtaining $\bar{r}$ from chemiluminescence images is detailed in Appendix \ref{app_order}. The oscillator distributions on the $\dot{\theta}_i-\psi_i$ plane for different states are shown in figure \ref{fig:Fig9_order_param}(a,b). Characteristics of oscillators from spatiotemporal images are depicted using lighter shade markers, while that from the model are illustrated using darker shade markers. As the heat release rate oscillators in experiments evolve in physical space, a comparison of oscillator properties in the phase space $\dot{\theta}_i-\psi_i$ allows us to gauge how closely the spatiotemporal synchronization of oscillators are captured by the low-dimensional dynamical mean-field model. Note that during the occurrence of combustion noise, the initial mean subtracted frequency distribution is $\dot{\theta}_i=\omega_i$ and $\dot{\theta}_i$ is suitably mean subtracted and normalized by the acoustic frequency $\Omega_0$ according to \eqref{Eq-3.9}.

For the bluff-body dump combustor, which shows a continuous transition to thermoacoustic instability through intermittency, the order parameter also shows a continuous and monotonous increase as $K$ is varied (figure \ref{fig:Fig9_order_param}a). For $K<0.75$ corresponding to the state of combustion noise (figure \ref{fig:Fig5_continuous}b), we observe the oscillators to have a broad distribution of $\dot{\theta}_i=\omega_i$ and $\psi_i$ (figure \ref{fig:Fig9_order_param}c). As the oscillators are desynchronized, $\bar{r}$ is close to zero. As $K$ is increased past $0.75$, intermittency appears in the system dynamics (figure \ref{fig:Fig9_order_param}d). This results in the appearance of the phase-synchronized cluster, which is small at first but grows in size with increasing $K$. At $K=0.95$, we notice that the frequency of oscillators is no longer broadly distributed and instead have a narrowband distribution around the mean frequency $\Omega_0$ (figure \ref{fig:Fig9_order_param}e). Due to a higher degree of synchronization among oscillators, $\bar{r}$ increases monotonously and continuously till the state of thermoacoustic instability. This can be observed at $K \approx 2$, at which $\dot{\theta}_i$ is close to zero, and all the oscillators fluctuate at the mean acoustic frequency, as can be observed from the sharp peak in $\mathcal{P}(\dot{\theta}_i)$ (figure \ref{fig:Fig9_order_param}f). Further, we notice that the oscillators are phase-locked with distribution in $\psi_i<|\pi/2|$. Accordingly, the order parameter is $\bar{r}=0.88$, implying global phase synchronization among the oscillators. 

In the case of the annular combustor, which undergoes a secondary bifurcation to thermoacoustic instability, $\bar{r}$ shows a discontinuous transition as $K$ is increased (figure \ref{fig:Fig9_order_param}b). As noted earlier, during the occurrence of combustion noise ($K<1.3$), the oscillators show broad frequency and phase distributions (figure \ref{fig:Fig9_order_param}g). The frequency distribution narrows close to zero during the state of intermittency (figure \ref{fig:Fig9_order_param}h). As $K$ is increased further, the state of low-amplitude limit cycle is reached, we observe a bimodal frequency distribution with a peak close to zero and another peak at $\dot{\theta}_i\approx -0.6$ (figure \ref{fig:Fig9_order_param}i). Upon increasing $K>1.6$, there is an abrupt jump in the value of $\bar{r}$ as the state of phase synchronization is reached, as confirmed from the sharp peak in $\mathcal{P}(\dot{\theta}_i$) (figure \ref{fig:Fig9_order_param}j). This is associated with the state of high-amplitude thermoacoustic instability in the annular combustor. Kindly also refer to supplementary movies S1-S8 to observe the time evolution of the oscillators.

\begin{figure}
\centering
\includegraphics[width=\linewidth]{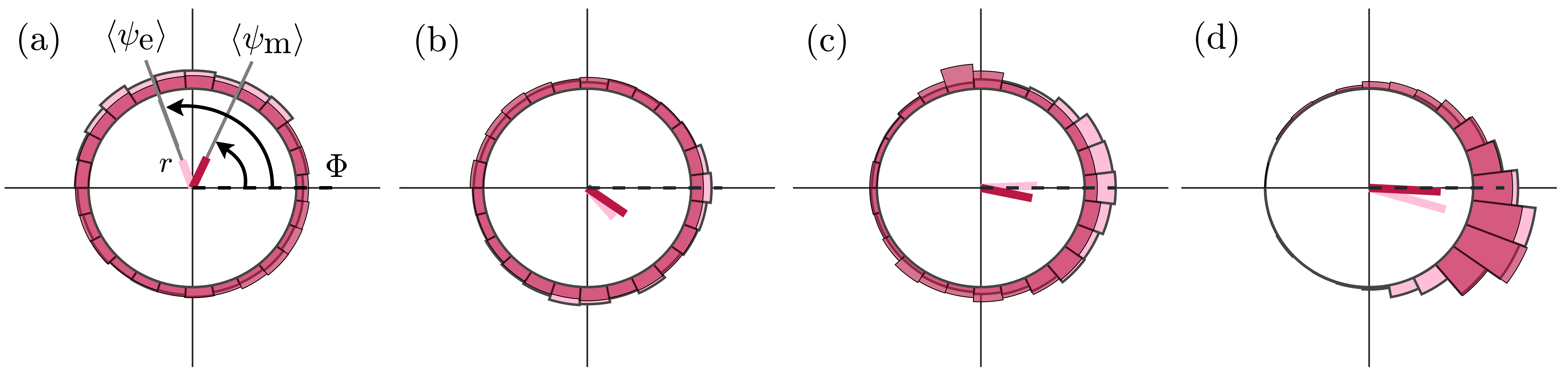}
\caption{Polar plot in the co-rotating frame showing the variation in the instantaneous distribution of relative phases ($\psi_i$) obtained from the experiments (dark shade) and model (light shade) during (a) combustion noise, (b) intermittency, (c) low-amplitude and (d) high-amplitude thermoacoustic instability. The representative states correspond to the points indicated in figure \ref{fig:Fig9_order_param}b observed during the abrupt \textit{first-order} explosive synchronization in the annular combustor. Averaged relative phase from the experiments and the model are indicated using $\langle \psi_\textrm{e} \rangle$ and $\langle \psi_\textrm{m} \rangle$ along with the Kuramoto order parameter ($r$), respectively. The dashed line indicates the reference phase of the acoustic pressure oscillations ($\Phi$). See also supplemental movies S1-S8.}
\label{fig:Fig10_phase_dist}
\end{figure}

To clarify the picture of synchronization, it is instructive to plot the distribution of relative phases $\psi_i=\theta_i-\Phi$ in polar coordinates. Since, by definition, the frequency of oscillators is centered around the acoustic frequency ($\Omega_0$), the frame of reference of the oscillators is co-rotating with respect to $\Omega_0$. Figure \ref{fig:Fig10_phase_dist} shows the instantaneous distribution of $\psi_i$ corresponding to four representative states observed in the annular combustor (cf. figure \ref{fig:Fig9_order_param}b). The experimental data are shown using a lighter shade, while the model results are shown using a darker shade. The instantaneous averaged relative phase obtained from the model $\langle\psi_\textrm{m}\rangle$ and the experiments $\langle\psi_\textrm{e}\rangle$ are also shown along with the respective Kuramoto order parameter ($\bar{r}$) obtained using \eqref{Eq-Kuramoto-Order} and \eqref{eq:appC1}. The plot shows a drastic change where initially asynchronous oscillators during the state of combustion noise become synchronous during high-amplitude thermoacoustic instability.

During the occurrence of combustion noise, the oscillators are uniformly distributed. The average phase also drifts with respect to $\Phi$, indicating desynchronization amongst the oscillators (see also movie S5). During the transition to thermoacoustic instability, we observe a clear shift from a uniform broadband distribution to a narrowband distribution, implying the emergence of synchronization among oscillators. The average phase remains locked to $\Phi$, and the exact phase difference never exceeds $\pm\pi/2$ (cf. movie S8). Finally, we see that this behavior is very well approximated by the model.

The stark contrast in the bifurcation behavior of the order parameter $\bar{r}$ shown in figure \ref{fig:Fig9_order_param}(a,b) highlights the difference in the characteristics of synchronization underlying the two bifurcations. Indeed, the continuous change in $\bar{r}$ exemplifies a \textit{second-order} synchronization phase-transition. On the other hand, the abrupt bifurcation in $\bar{r}$ embodies a \textit{first-order} phase transition and is more appropriately referred to as \textit{explosive} synchronization \citep{strogatz2000kuramoto,pazo2005thermodynamic,leyva2013explosive,kuehn2021universal}. 

Whether the synchronization transition will be continuous or explosive is crucially contingent on the initial frequency distribution. It is well-known that the second-order continuous transitions appear in the standard Kuramoto model whenever the frequency distribution of phase oscillators is symmetric and unimodal \citep{kuramoto1975self, strogatz2000kuramoto}. This is due to the presence of a clear peak in the distribution, which ensures that upon increasing coupling strength, a large cluster of oscillators gets synchronized around the peak in the distribution. Further increase in coupling strength leads to entrainment of drifting oscillators to the large cluster resulting in the gradual increase in the size of the coherent cluster \citep{strogatz2000kuramoto, basnarkov2007phase}. 

The picture becomes complicated when the frequency distribution undergoes symmetry breaking to non-unimodal and asymmetric distribution in non-standard extensions of the Kuramoto model \citep{zhou2015explosive, terada2017nonstandard, de2020nonmonotonic}. For instance, in the case of bimodal distribution, the appearance of two frequency peaks means that an increase in the coupling strength leads to the entrainment of oscillators distributed around the two frequency peaks. When the coupling strength becomes too large, these peaks and the two clusters coalesce abruptly, leading to non-standard, first-order explosive synchronization \citep{terada2017nonstandard, zhang2020synchronization}. Similar observation has been made when the frequency distribution is flat \citep{basnarkov2007phase, pietras2018first} or asymmetric unimodal \citep{zhou2015explosive, de2020nonmonotonic}. While the above-mentioned studies are important steps in understanding second-order and first-order transitions, a clear resolution is missing as yet.

Here also, the key to discerning the reason behind second-order and first-order transitions lies in the characteristics of the frequency distribution. Evidently, the distributions $g(\omega)$ obtained from the three experiments are non-standard and asymmetric, as can be observed in figure \ref{fig:Fig4_freq_var}. The distributions are more clearly shown in figure \ref{fig:Fig9_order_param}(c,g). The distribution $g(\omega)$ for the bluff-body stabilized combustor is multimodal, where the peaks are centered very close to the frequency of acoustic oscillations (figure \ref{fig:Fig9_order_param}c). Now, as the coupling strength increases, oscillators at these frequencies get entrained, and a single peak is established in the frequency distribution (cf. figure \ref{fig:Fig9_order_param}e). Further increase in the coupling strength leads to a gradual increase in the size of the largest entrained cluster. Hence, we observe a continuous, second-order synchronization transition. In contrast, for the annular combustor, the distribution $g(\omega)$ is initially asymmetric with a peak that is comparatively farther away from the frequency of acoustic fluctuations (cf. figure \ref{fig:Fig9_order_param}g). Now, as the coupling strength is increased, a secondary peak becomes clearly visible (cf. figure \ref{fig:Fig9_order_param}i). Thus, oscillators are entrained around two different clusters associated with the two peaks. An increase in $K$ beyond a critical value leads to an abrupt coalescence of these two clusters, resulting in the first-order explosive synchronization. 

To summarize, we have seen that although the model is dynamical with no spatial input, it captures the characteristics of spatiotemporal synchronization patterns observed in experiments very well while also predicting the nature of bifurcation to limit cycle oscillations--a feature that has yet to be captured in other thermoacoustic models. Thus, the above results strongly suggest the usefulness of the proposed mean-field model for analyzing the thermoacoustic transitions in turbulent combustion systems. 

It is worth mentioning here that explosive synchronization has been reported in power-grids \citep{motter2013spontaneous}, neurological activity \citep{kim2016functional}, chemical reactions \citep{kumar2015experimental, cualuguaru2020first}. Our study is the first experimental evidence of explosive synchronization in a strongly-coupled fluid dynamical system.

\section{Concluding remarks}
\label{Sec7_Conclusion}
To summarise, we have presented a mean-field synchronization model for predicting transitions to thermoacoustic instability in turbulent combustion systems. We have seen that thermoacoustic transitions can be continuous or abrupt and arise through spatiotemporal synchronization. To explain such a rich dynamical behavior, we assume that the turbulent flame comprises an ensemble of phase oscillators evolving under the influence of mean-field interactions and acoustic feedback. These interactions encode the nonlinearities in the flame response subjected to acoustic and turbulent fluctuations. 

We showed that the mean-field model captures continuous and abrupt transitions observed in three distinct (bluff-body stabilized, swirl-stabilized, and annular) combustor configurations. These transitions are captured by the model by taking the heat release rate spectrum during the stable operation as the only input. Further, the model captures the characteristics such as time series, PDF, and spectrum of the different states--combustion noise, intermittency, limit cycle oscillations--en route to the state of thermoacoustic instability in these systems. We then estimated the relationship between experimental and model parameters using a gradient descent algorithm. In all three combustors, we find that the coupling strength is a linear function of the equivalence ratio, indicating that a change in the control parameter leads to an increase in the coupling strength of the phase oscillators. Such a relationship highlights the interpretability of the model: a change in an experimental control parameter leads to an increase in the coupling strength of the phase oscillators, promoting synchronization and limit cycle oscillations. 

Importantly, we show that our modelling approach naturally provides an explanation of spatiotemporal synchronization and pattern formation observed in turbulent thermoacoustic systems. We showed that the model closely captures the statistical behavior of spatial desynchronization, chimera, and global phase synchronization underlying the transitions. Our results strongly indicate that continuous and abrupt thermoacoustic transitions are associated with synchronization transition of second-order and first-order, respectively. This observation of disparate phase transitions is further rationalized based on the frequency spectrum of the phase oscillators. We observe the appearance of a unimodal peak around which all the oscillators get entrained, giving rise to second-order transition. On the other hand, the first-order explosive transition is associated with the appearance of a bimodal distribution where two synchronized clusters of oscillators get entrained. An increase in the coupling strength beyond a critical point results in a sudden, abrupt coalescence into one large synchronized cluster. 

Thus, the proposed mean-field model not only explains distinct types of bifurcation to limit cycle oscillations in disparate systems but also does so in a consistent manner based on the paradigm of synchronization without the need for disparate modeling approaches. Secondary effects such as multi-modal interactions, the effect of turbulence, and stochastic forcing on the phase synchronization of oscillators are left for future studies. 

Our study provides a fresh perspective concerning the connection between synchronization and thermoacoustic transitions. In the broader context of nonlinear dynamics, our results provide valuable experimental evidence of explosive synchronization in a fluid dynamical setting and may help in resolving the issues surrounding the resolution of second-order and first-order synchronization in non-standard Kuramoto models. Finally, our approach opens further avenues for the modeling of related fluid dynamical systems such as aeroacoustic and flow-structure interactions where spatiotemporal interactions lead to rich dynamical phenomena.

\backsection[Acknowledgements]{Samarjeet, Amitesh, and Jayesh gratefully acknowledge the Ministry of Human Resource Development (MHRD) for Ph.D. funding through the Half-Time Research Assistantship (HTRA). We also acknowledge Induja Pavithran, Manikandan Raghunathan, Midhun Raghunathan, Thilagaraj S., and Anand Selvam for their help in performing experiments. We also express our gratitude to Ankit Sahay, Sneha Srikanth, and Alan J. Varghese for fruitful discussions on the model.}

\backsection[Funding]{R. I. Sujith is grateful for the funding from the Institute of Eminence (IoE) initiative of IIT Madras (SB/2021/0845/AE/MHRD/002696) and the Office of Naval Research Global (Grant No. N62909-18-1-2061; Funder ID: 10.13039/100007297). S. Chaudhuri acknowledges support from the Natural Sciences and Engineering Research Council of Canada Discovery Grant (RGPIN-2021-02676).}

\backsection[Declaration of interests]{The authors report no conflict of interest.}

\backsection[Data availability statement]{The data that support the findings of this study are available upon reasonable request from the corresponding authors.}

\backsection[Author ORCID]{
\begin{enumerate}
\item[] \orcidlink{0000-0002-4372-2711} Samarjeet Singh \href{https://orcid.org/0000-0002-4372-2711}{https://orcid.org/0000-0002-4372-2711}
\item[] \orcidlink{0000-0002-8192-5448} Amitesh Roy \href{https://orcid.org/0000-0002-8192-5448}{https://orcid.org/0000-0002-8192-5448}
\item[] \orcidlink{0000-0003-3502-9009} Jayesh M. Dhadphale \href{https://orcid.org/0000-0003-3502-9009}{https://orcid.org/0000-0003-3502-9009}
\item[] \orcidlink{0000-0003-4109-8633} Swetaprovo Chaudhuri \href{https://orcid.org/0000-0003-4109-8633}{https://orcid.org/0000-0003-4109-8633}%

\item[] \orcidlink{https://orcid.org/0000-0002-0791-7896
} R. I. Sujith \href{https://orcid.org/0000-0002-0791-7896
}{https://orcid.org/0000-0002-0791-7896}
\end{enumerate}
}

\backsection[Author contributions]{All the authors contributed in formulating the problem and writing the paper.}

\appendix

\section{Numerical procedure for sampling oscillator frequency distribution from experimental data}
\label{app_freq}

The density of oscillators in frequency domain $g(\omega)$ is obtained from $\hat{\dot{q}}(\omega)$, i.e., Fourier transform of the time series of $\dot{q}'(t)$ during the occurrence of combustion noise. The $\hat{\dot{q}}(\omega)$ is available for discrete frequencies, $\{\omega_1, \omega_2,...,\omega_{N_{F}}\}$, where $\omega_1=0$ and $\omega_{N_{F}}$ is the maximum frequency. The $\hat{\dot{q}}(\omega)$ is normalized to obtain $g(\omega)= |\hat{\dot{q}}(\omega)|/\int_0^{\omega_{N_F}} |\hat{\dot{q}}(\omega')|d\omega'$. The normalization ensures $\int_{0}^{\omega_{N_F}} g(\omega)d\omega=1$. The integration is performed numerically with the trapezoidal rule. This procedure gives $g(\omega)$ at discrete frequencies, which is used for sampling the frequency for each of the $N$ phase oscillators. 
  
To obtain samples from arbitrary discrete distribution $g(\omega)$, we use the uniform distribution $U(x)$ with support over $x\in[0,C_{N_F}]$, where $C_k=\sum_{i=1}^{k} g(\omega_k)$, i.e. for $k=N_F$ we get $C_{N_F}$ as cumulative sum over all the discrete $g(\omega)$ values. The $N$ data points $\{x_1, x_2,...,x_N\}$ are sampled from $U$. The frequency of $i^{\textrm{th}}$ oscillator is then obtained as 
\begin{equation} \label{eqappa1}
\omega_{s,i}=(1-\alpha)\omega_k + \alpha\omega_{k+1},
\end{equation} 
where, 
\begin{equation} 
\alpha=(x_i-C_k)/(C_{k+1}-C_k),\notag
\end{equation}
and $k$ satisfies $C_k\leq x_i<C_{k+1}$. 
This procedure samples the frequency of $N$ phase oscillators according to the normalized distribution $g(\omega )$ obtained from the experimental data.

\section{Sensitivity analysis of parameter estimation}
\label{app_loss}
\begin{figure}
\centering
\includegraphics[width=0.7\textwidth]{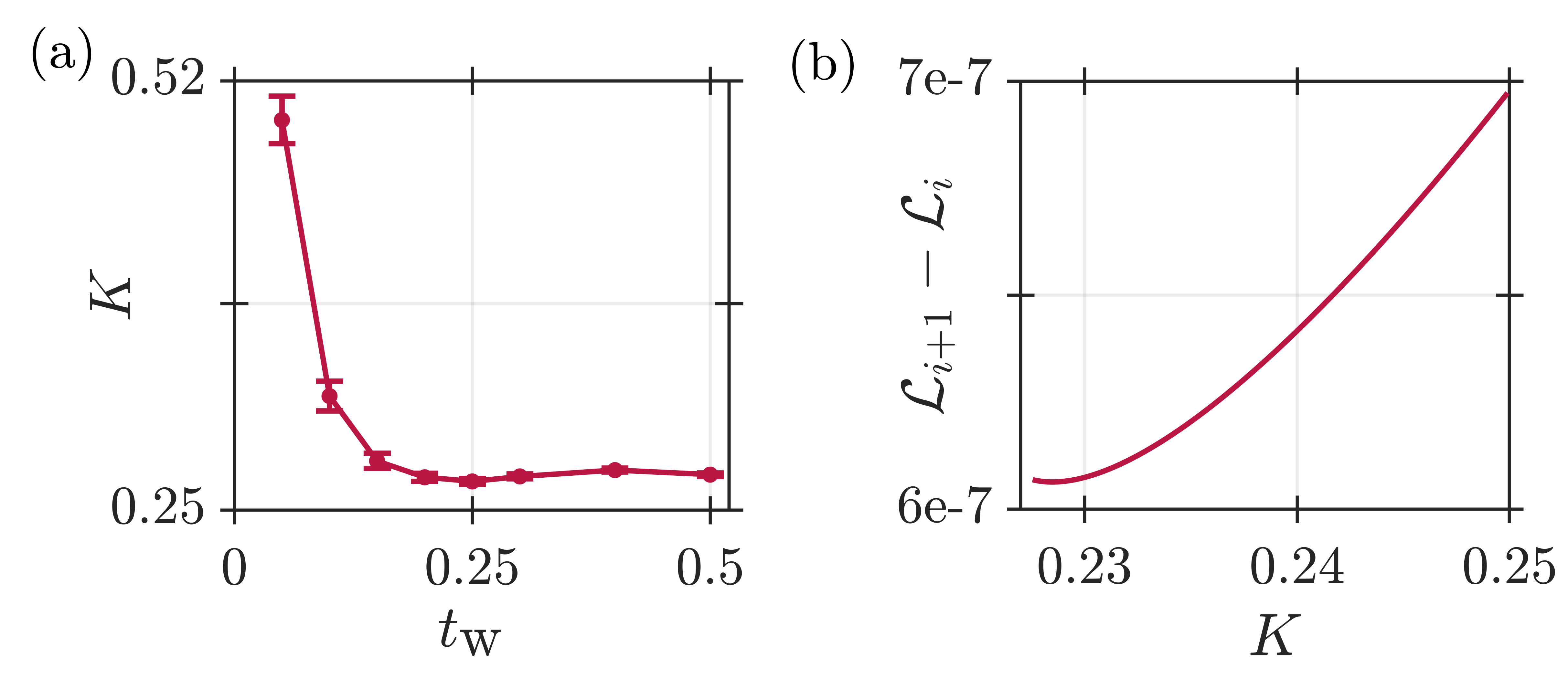}
\caption{Convergence of parameter optimisation algorithm. Panel (a) shows the variation in the estimated coupling strength ($K$) as a function of the size of the time window ($t_\textrm{w}$) used for optimizing the loss function $\mathcal{L}(\mathbb{P})$ for a fixed number of 1000 iterations. (b) Realisation of the minimisation scheme showing a change in the difference $\mathcal{L}_{i+1}-\mathcal{L}_i$ when the optimisation is performed with an initial value of $K=0.25$ over 1000 iterations during the state of combustion noise in bluff-body stabilized combustor.}
\label{fig:Fig11_convg}
\end{figure}

Estimating the parameter $\mathbb{P}$ by minimizing the error in $\textbf{X}_{\textrm{m}}$ and $\textbf{X}_{\textrm{exp}}$ is numerically expensive. So, only a portion ($t_{\textrm{w}}$) of the entire time series $\textbf{X}_{\textrm{exp}}$ is used for parameter optimisation. To ensure convergence in the estimate of the model parameter, we vary the length of the time series $t_\textrm{w}$ used for optimisation. For each window size, the optimisation is performed for 1000 iterations to determine the value of $K$. Then the window is moved across the entire length of the time series to obtain a distribution of $K$. The standard deviation of this distribution is then used to obtain the error bar. Figure \ref{fig:Fig11_convg}(a) shows the convergence of the model control parameter ($K$) as a function of the time window ($t_\textrm{w}$) used for performing optimisation according to \eqref{Eq-4.5}. We find that the error in estimation $K$ is quite low. We notice that the value of $K$ reaches a constant value after a window of size $t_w\approx0.25$ s. Thus, we use $t_w=0.5$ s for optimizing parameter across all data-sets.

Figure \ref{fig:Fig11_convg}(b) shows a typical realisation of an optimisation performed with the initial guess of $K=0.25$ and time window of size $t_w=0.4$ s.pixels The minimisation is performed over 1000 iterations. The plot shows the manner in which $\mathcal{L}_{i+1}-\mathcal{L}_i$ reduces with an increasing number of iterations. The difference in the value of the loss function is of the order of $~10^{-7}$. The optimized value of $K$ corresponds to a minima in $\mathcal{L}_{i+1}-\mathcal{L}_i$.

\section{Extracting order parameter from spatiotemporal experimental data}
\label{app_order}

To extract the order parameter from a spatiotemporal data, it is important to process $\dot{q}^\prime(x,y,t)$ properly. First, to reduce the effect of noise in the spatiotemporal data, we coarse-grain chemiluminescence images over $8\times8$ and $6\times6$ pixels for the bluff-body stabilized combustor and annular combustor. We then normalize the time series during various states of operation by the amplitude $\dot{q}^\prime$ during limit cycle oscillations. The resulting signal at each coarse-grained location depicts a transition from a low amplitude chaotic state to limit cycle oscillations of amplitude when the control parameter is varied. 

\citet{popovych2005phase} and \citet{bick2011chaos} showed that the collective behavior of oscillators with distributed frequencies yields chaotic behavior. Following the same approach, we assume that the heat release rate fluctuations measured at each coarse-grained location are a result of a set of limit-cycle oscillators. In other words, we assume that $k^\textrm{th}$ pixel comprises $n_k$ number of limit cycle oscillators. To simplify calculations, we assume that $n_k=n$ for all the pixels. Let the phase for $j^{\textrm{th}}$ oscillator at $k^\textrm{th}$ pixel is $\varphi_{kj}$, where $j = 1,..,n$. Therefore, the complex order parameter for $k^\textrm{th}$ pixel is expressed as: $r_k e^{i \theta_k} = 1/n \sum_{j = 1}^{n} e^{i \varphi_{kj}}$ \citep{strogatz2000kuramoto}, where $r_k(t)$ and $\psi_k(t)$ can be simply obtained from the absolute value and argument of the Hilbert transform of $\dot{q}^\prime_k(t)$. Consequently, the order parameter can be defined as:
\begin{equation}
  \bar{r}e^{i\langle\theta\rangle} = \left\langle\frac{1}{N_p} \sum_{k=1}^{N_p}r_ke^{i \psi_k}\right\rangle_t,
  \label{eq:appC1}
\end{equation}
where successive averaging operations were taken over $N_p$ pixels in each image and the total number of chemiluminescence images in the time series. The value of $\bar{r}$ so determined is then used in figure \ref{fig:Fig9_order_param}a,b.

\bibliographystyle{jfm}
\bibliography{references}

\begin{thebibliography}{90}
\expandafter\ifx\csname natexlab\endcsname\relax\def\natexlab#1{#1}\fi
\def\au#1{#1} \def\ed#1{#1} \def\yr#1{#1}\def\at#1{#1}\def\jt#1{\textit{#1}}
  \def\bt#1{#1}\def\bvol#1{\textbf{#1}} \def\vol#1{#1} \def\pg#1{#1}
  \def\publ#1{#1}\def\arxiv#1{#1}\def\org#1{#1}\def\st#1{\textit{#1}}

\bibitem[Abrams(2006)]{abrams2006two}
{\sc \au{Abrams, D.~M.}} \yr{2006} {\em Two coupled oscillator models: the
  {M}illennium {B}ridge and the chimera state \rm{(PhD Thesis)}\/}.
  \publ{Cornell University, New York}.

\bibitem[Abrams \& Strogatz(2004)]{abrams2004chimera}
{\sc \au{Abrams, D.~M.} \& \au{Strogatz, S.~H.}} \yr{2004}  \at{Chimera states
  for coupled oscillators}.  \jt{Phys. Rev. Appl.}  \bvol{93}~(17),
  \pg{174102}.

\bibitem[Agharkar {\em et~al.\/}(2013)Agharkar, Subramanian, Kaisare \&
  Sujith]{agharkar2013thermoacoustic}
{\sc \au{Agharkar, P.}, \au{Subramanian, P.}, \au{Kaisare, N.S.} \& \au{Sujith,
  R.~I.}} \yr{2013}  \at{Thermoacoustic instabilities in a ducted premixed
  flame: reduced-order models and control}.  \jt{Combust. Sci. Technol.}
  \bvol{185}~(6),  \pg{920--942}.

\bibitem[Ananthkrishnan {\em et~al.\/}(2005)Ananthkrishnan, Deo \&
  Culick]{ananthkrishnan2005reduced}
{\sc \au{Ananthkrishnan, N.}, \au{Deo, S.} \& \au{Culick, F. E.~C.}} \yr{2005}
  \at{Reduced-order modeling and dynamics of nonlinear acoustic waves in a
  combustion chamber}.  \jt{Combust. Sci. Technol.}  \bvol{177}~(2),
  \pg{221--248}.

\bibitem[Ananthkrishnan {\em et~al.\/}(1998)Ananthkrishnan, Sudhakar, Sudershan
  \& Agarwal]{ananthkrishnan1998application}
{\sc \au{Ananthkrishnan, N.}, \au{Sudhakar, K.}, \au{Sudershan, S.} \&
  \au{Agarwal, A.}} \yr{1998}  \at{Application of secondary bifurcations to
  large-amplitude limit cycles in mechanical systems}.  \jt{J. Sound Vib.}
  \bvol{215}~(1),  \pg{183--188}.

\bibitem[Balanov {\em et~al.\/}(2008)Balanov, Janson, Postnov \&
  Sosnovtseva]{balanov2008synchronization}
{\sc \au{Balanov, A.}, \au{Janson, N.}, \au{Postnov, D.} \& \au{Sosnovtseva,
  O.}} \yr{2008} {\em Synchronization: {F}rom simple to complex\/}.
  \publ{Springer Science \& Business Media}.

\bibitem[Balasubramanian \& Sujith(2008)]{balasubramanian2008thermoacoustic}
{\sc \au{Balasubramanian, K.} \& \au{Sujith, R.~I.}} \yr{2008}
  \at{Thermoacoustic instability in a {R}ijke tube: Non-normality and
  nonlinearity}.  \jt{Phys. Fluids}  \bvol{20}~(4),  \pg{044103}.

\bibitem[Basnarkov \& Urumov(2007)]{basnarkov2007phase}
{\sc \au{Basnarkov, L.} \& \au{Urumov, V.}} \yr{2007}  \at{Phase transitions in
  the {K}uramoto model}.  \jt{Phys. Rev. E}  \bvol{76}~(5),  \pg{057201}.

\bibitem[Baydin {\em et~al.\/}(2018)Baydin, Pearlmutter, Radul \&
  Siskind]{baydin2018automatic}
{\sc \au{Baydin, A.~G.}, \au{Pearlmutter, B.A.}, \au{Radul, A.~A.} \&
  \au{Siskind, J.~M.}} \yr{2018}  \at{Automatic differentiation in machine
  learning: a survey}.  \jt{J. Mach. Learn. Res.}  \bvol{18}.

\bibitem[Bhavi {\em et~al.\/}(2022)Bhavi, Pavithran, Roy \&
  Sujith]{bhavi2022abrupt}
{\sc \au{Bhavi, R.~S.}, \au{Pavithran, I.}, \au{Roy, A.} \& \au{Sujith, R.~I.}}
  \yr{2022}  \at{Abrupt transitions in turbulent thermoacoustic systems}.
  \jt{arXiv:2204.01342} .

\bibitem[Bick {\em et~al.\/}(2011)Bick, Timme, Rathlev \&
  Ashwin]{bick2011chaos}
{\sc \au{Bick, C.}, \au{Timme, M.and~Paulikat, D.}, \au{Rathlev, D.} \&
  \au{Ashwin, P.}} \yr{2011}  \at{Chaos in symmetric phase oscillator
  networks}.  \jt{Phys. Rev. Lett.}  \bvol{107}~(24),  \pg{244101}.

\bibitem[Bonciolini {\em et~al.\/}(2017)Bonciolini, Boujo \&
  Noiray]{bonciolini2017output}
{\sc \au{Bonciolini, G.}, \au{Boujo, E.} \& \au{Noiray, N.}} \yr{2017}
  \at{Output-only parameter identification of a colored-noise-driven
  {V}an-der-{P}ol oscillator: thermoacoustic instabilities as an example}.
  \jt{Phys. Rev. E}  \bvol{95}~(6),  \pg{062217}.

\bibitem[Bonciolini {\em et~al.\/}(2021)Bonciolini, Faure-Beaulieu, Bourquard
  \& Noiray]{bonciolini2021low}
{\sc \au{Bonciolini, G.}, \au{Faure-Beaulieu, A.}, \au{Bourquard, C.} \&
  \au{Noiray, N.}} \yr{2021}  \at{Low order modelling of thermoacoustic
  instabilities and intermittency: Flame response delay and nonlinearity}.
  \jt{Combust. Flame}  \bvol{226},  \pg{396--411}.

\bibitem[Boyd {\em et~al.\/}(2004)Boyd, Boyd \& Vandenberghe]{boyd2004convex}
{\sc \au{Boyd, S.}, \au{Boyd, S.~P.} \& \au{Vandenberghe, L.}} \yr{2004} {\em
  Convex optimization\/}.  \publ{Cambridge university press, New York}.

\bibitem[C{\u{a}}lug{\u{a}}ru {\em et~al.\/}(2020)C{\u{a}}lug{\u{a}}ru, Totz,
  Martens \& Engel]{cualuguaru2020first}
{\sc \au{C{\u{a}}lug{\u{a}}ru, D.}, \au{Totz, J.~F.}, \au{Martens, E.~A.} \&
  \au{Engel, H.}} \yr{2020}  \at{First-order synchronization transition in a
  large population of strongly coupled relaxation oscillators}.  \jt{Sci. Adv.}
   \bvol{6}~(39),  \pg{eabb2637}.

\bibitem[Candel {\em et~al.\/}(2009)Candel, Durox, Ducruix, Birbaud, Noiray \&
  Schuller]{candel2009flame}
{\sc \au{Candel, S.~M.}, \au{Durox, D.}, \au{Ducruix, S.}, \au{Birbaud, A.~L.},
  \au{Noiray, N.} \& \au{Schuller, T.}} \yr{2009}  \at{Flame dynamics and
  combustion noise: progress and challenges}.  \jt{Int. J. Aeroacoustics}
  \bvol{8}~(1),  \pg{1--56}.

\bibitem[Chu(1965)]{chu1965energy}
{\sc \au{Chu, B.~T.}} \yr{1965}  \at{On the energy transfer to small
  disturbances in fluid flow ({P}art {I})}.  \jt{Acta Mech.}  \bvol{1}~(3),
  \pg{215--234}.

\bibitem[Culick(2006)]{culick2006unsteady}
{\sc \au{Culick, F. E.~C.}} \yr{2006}  \bt{Unsteady motions in combustion
  chambers for propulsion systems}. {\em Tech. Rep.\/}.  \org{NATO, AGARDograph
  AG-AVT-039}.

\bibitem[Dowling(1995)]{dowling1995calculation}
{\sc \au{Dowling, A.~P.}} \yr{1995}  \at{The calculation of thermoacoustic
  oscillations}.  \jt{J. Sound Vib.}  \bvol{180}~(4),  \pg{557--581}.

\bibitem[Dutta {\em et~al.\/}(2019)Dutta, Ramachandran \&
  Chaudhuri]{dutta2019investigating}
{\sc \au{Dutta, A.~K.}, \au{Ramachandran, G.} \& \au{Chaudhuri, S.}} \yr{2019}
  \at{Investigating thermoacoustic instability mitigation dynamics with a
  {K}uramoto model for flamelet oscillators}.  \jt{Phys. Rev. E}
  \bvol{99}~(3),  \pg{032215}.

\bibitem[Eckhardt {\em et~al.\/}(2007)Eckhardt, Ott, Strogatz, Abrams \&
  McRobie]{eckhardt2007modeling}
{\sc \au{Eckhardt, B.}, \au{Ott, E.}, \au{Strogatz, S.~H.}, \au{Abrams, D.~M.}
  \& \au{McRobie, A.}} \yr{2007}  \at{Modeling walker synchronization on the
  {M}illennium {B}ridge}.  \jt{Phys. Rev. E}  \bvol{75}~(2),  \pg{021110}.

\bibitem[George {\em et~al.\/}(2018)George, Unni, Raghunathan \&
  Sujith]{george2018pattern}
{\sc \au{George, N.~B.}, \au{Unni, V.~R.}, \au{Raghunathan, M.} \& \au{Sujith,
  R.~I.}} \yr{2018}  \at{Pattern formation during transition from combustion
  noise to thermoacoustic instability via intermittency}.  \jt{J. Fluid Mech.}
  \bvol{849},  \pg{615--644}.

\bibitem[Ghirardo \& Juniper(2013)]{ghirardo2013azimuthal}
{\sc \au{Ghirardo, G.} \& \au{Juniper, M.~P.}} \yr{2013}  \at{Azimuthal
  instabilities in annular combustors: standing and spinning modes}.  \jt{Proc.
  R. Soc. A}  \bvol{469}~(2157),  \pg{20130232}.

\bibitem[Godavarthi {\em et~al.\/}(2020)Godavarthi, Kasthuri, Mondal, Sujith,
  Marwan \& Kurths]{godavarthi2020synchronization}
{\sc \au{Godavarthi, V.}, \au{Kasthuri, P.}, \au{Mondal, S.}, \au{Sujith,
  R.~I.}, \au{Marwan, N.} \& \au{Kurths, J.}} \yr{2020}  \at{Synchronization
  transition from chaos to limit cycle oscillations when a locally coupled
  chaotic oscillator grid is coupled globally to another chaotic oscillator}.
  \jt{Chaos}  \bvol{30}~(3),  \pg{033121}.

\bibitem[Gotoda {\em et~al.\/}(2011)Gotoda, Nikimoto, Miyano \&
  Tachibana]{gotoda2011dynamic}
{\sc \au{Gotoda, H.}, \au{Nikimoto, H.}, \au{Miyano, T.} \& \au{Tachibana, S.}}
  \yr{2011}  \at{Dynamic properties of combustion instability in a lean
  premixed gas-turbine combustor}.  \jt{Chaos}  \bvol{21}~(1),  \pg{013124}.

\bibitem[Guan {\em et~al.\/}(2019)Guan, Li, Ahn \& Kim]{guan2019chaos}
{\sc \au{Guan, Y.}, \au{Li, L. K.~B.}, \au{Ahn, B.} \& \au{Kim, K.~T.}}
  \yr{2019}  \at{Chaos, synchronization, and desynchronization in a
  liquid-fueled diffusion-flame combustor with an intrinsic hydrodynamic mode}.
   \jt{Chaos}  \bvol{29}~(5),  \pg{053124}.

\bibitem[Hashimoto {\em et~al.\/}(2019)Hashimoto, Shibuya \&
  Ohmichi]{hashimoto2019spatiotemporal}
{\sc \au{Hashimoto, T.}, \au{Shibuya, H.and~Gotoda, H.} \& \au{Ohmichi,
  Y.and~Matsuyama, S.}} \yr{2019}  \at{Spatiotemporal dynamics and early
  detection of thermoacoustic combustion instability in a model rocket
  combustor}.  \jt{Phys. Rev. E}  \bvol{99}~(3),  \pg{032208}.

\bibitem[Kheirkhah {\em et~al.\/}(2017)Kheirkhah, Cirtwill, Saini, Venkatesan
  \& Steinberg]{kheirkhah2017dynamics}
{\sc \au{Kheirkhah, S.}, \au{Cirtwill, J. D.~M.}, \au{Saini, P.},
  \au{Venkatesan, K.} \& \au{Steinberg, A.~M.}} \yr{2017}  \at{Dynamics and
  mechanisms of pressure, heat release rate, and fuel spray coupling during
  intermittent thermoacoustic oscillations in a model aeronautical combustor at
  elevated pressure}.  \jt{Combust. Flame}  \bvol{185},  \pg{319--334}.

\bibitem[Kim {\em et~al.\/}(2016)Kim, Mashour, Moraes, Vanini, Tarnal, Janke,
  Hudetz \& Lee]{kim2016functional}
{\sc \au{Kim, M.}, \au{Mashour, G.~A.}, \au{Moraes, S.~B.}, \au{Vanini, G.},
  \au{Tarnal, V.}, \au{Janke, E.}, \au{Hudetz, A.~G.} \& \au{Lee, U.}}
  \yr{2016}  \at{Functional and topological conditions for explosive
  synchronization develop in human brain networks with the onset of
  anesthetic-induced unconsciousness}.  \jt{Front. Comput. Neurosci.}
  \bvol{10},  \pg{1}.

\bibitem[Krylov \& Bogolyubov(1947)]{krylov1947introduction}
{\sc \au{Krylov, N.~M.} \& \au{Bogolyubov, N.~N.}} \yr{1947} {\em Introduction
  to nonlinear mechanics\/}.  \publ{Princeton University Press}.

\bibitem[Kuehn \& Bick(2021)]{kuehn2021universal}
{\sc \au{Kuehn, C.} \& \au{Bick, C.}} \yr{2021}  \at{A universal route to
  explosive phenomena}.  \jt{Sci. Adv.}  \bvol{7}~(16),  \pg{eabe3824}.

\bibitem[Kumar {\em et~al.\/}(2015)Kumar, Verma, Parmananda \&
  Boccaletti]{kumar2015experimental}
{\sc \au{Kumar, P.}, \au{Verma, D.~K.}, \au{Parmananda, P.} \& \au{Boccaletti,
  S.}} \yr{2015}  \at{Experimental evidence of explosive synchronization in
  mercury beating-heart oscillators}.  \jt{Phys. Rev. E}  \bvol{91}~(6),
  \pg{062909}.

\bibitem[Kuramoto(1975)]{kuramoto1975self}
{\sc \au{Kuramoto, Y.}} \yr{1975} Self-entrainment of a population of coupled
  non-linear oscillators.  \bt{In {\em International symposium on mathematical
  problems in theoretical physics. Lecture Notes in Physics\/}},  \pg{pp.
  420--422}. Springer.

\bibitem[Kuramoto(2003)]{kuramoto2003chemical}
{\sc \au{Kuramoto, Y.}} \yr{2003} {\em Chemical oscillations, waves, and
  turbulence\/}, ,  \vol{vol.~19}.  \publ{Springer Science \& Business Media.}

\bibitem[Kuramoto \& Battogtokh(2002)]{kuramoto2002coexistence}
{\sc \au{Kuramoto, Y.} \& \au{Battogtokh, D.}} \yr{2002}  \at{Coexistence of
  coherence and incoherence in nonlocally coupled phase oscillators}.
  \jt{arXiv preprint cond-mat/0210694} .

\bibitem[Laera {\em et~al.\/}(2017)Laera, Schuller, Prieur, Durox, Camporeale
  \& Candel]{laera2017flame}
{\sc \au{Laera, D.}, \au{Schuller, T.}, \au{Prieur, K.}, \au{Durox, D.},
  \au{Camporeale, S.~M.} \& \au{Candel, S.~M.}} \yr{2017}  \at{Flame describing
  function analysis of spinning and standing modes in an annular combustor and
  comparison with experiments}.  \jt{Combust. Flame}  \bvol{184},
  \pg{136--152}.

\bibitem[Lee {\em et~al.\/}(2020)Lee, Guan, Gupta \& Li]{lee2020input}
{\sc \au{Lee, M.}, \au{Guan, Y.}, \au{Gupta, V.} \& \au{Li, L. K.~B.}}
  \yr{2020}  \at{Input-output system identification of a thermoacoustic
  oscillator near a {H}opf bifurcation using only fixed-point data}.  \jt{Phys.
  Rev. E}  \bvol{101}~(1),  \pg{013102}.

\bibitem[Lee {\em et~al.\/}(2021)Lee, Kim, Gupta \& Li]{lee2021system}
{\sc \au{Lee, M.}, \au{Kim, K.~T.}, \au{Gupta, V.} \& \au{Li, L. K.~B.}}
  \yr{2021}  \at{System identification and early warning detection of
  thermoacoustic oscillations in a turbulent combustor using its noise-induced
  dynamics}.  \jt{Proc. Combust. Inst.}  \bvol{38}~(4),  \pg{6025--6033}.

\bibitem[Leyva {\em et~al.\/}(2013)Leyva, Navas, Sendi{\~n}a-Nadal, Almendral,
  Buld{\'u}, Zanin, Papo \& Boccaletti]{leyva2013explosive}
{\sc \au{Leyva, I.}, \au{Navas, A.}, \au{Sendi{\~n}a-Nadal, I.}, \au{Almendral,
  J.~A.}, \au{Buld{\'u}, J.~M.}, \au{Zanin, M.}, \au{Papo, D.} \&
  \au{Boccaletti, S.}} \yr{2013}  \at{Explosive transitions to synchronization
  in networks of phase oscillators}.  \jt{Sci. Rep.}  \bvol{3}~(1),  \pg{1--5}.

\bibitem[Lieuwen(2002)]{lieuwen2002experimental}
{\sc \au{Lieuwen, T.~C.}} \yr{2002}  \at{Experimental investigation of
  limit-cycle oscillations in an unstable gas turbine combustor}.  \jt{J.
  Propuls. Power}  \bvol{18}~(1),  \pg{61--67}.

\bibitem[Lieuwen(2003)]{lieuwen2003modeling}
{\sc \au{Lieuwen, T.~C.}} \yr{2003}  \at{Modeling premixed combustion-acoustic
  wave interactions: A review}.  \jt{J. Propuls. Power}  \bvol{19}~(5),
  \pg{765--781}.

\bibitem[Lieuwen(2012)]{lieuwen2012unsteady}
{\sc \au{Lieuwen, T.~C.}} \yr{2012} {\em Unsteady combustor physics\/}.
  \publ{Cambridge University Press}.

\bibitem[Lieuwen \& Yang(2005)]{lieuwen2005combustion}
{\sc \au{Lieuwen, T.~C.} \& \au{Yang, V.}} \yr{2005} {\em Combustion
  instabilities in gas turbine engines\/}.  \publ{AIAA}.

\bibitem[Lores \& Zinn(1973)]{lores1973nonlinear}
{\sc \au{Lores, M.~E.} \& \au{Zinn, B.~T.}} \yr{1973}  \at{Nonlinear
  longitudinal combustion instability in rocket motors}.  \jt{Combust. Sci.
  Technol.}  \bvol{7}~(6),  \pg{245--256}.

\bibitem[Matveev \& Culick(2003)]{matveev2003model}
{\sc \au{Matveev, K.~I.} \& \au{Culick, F. E.~C.}} \yr{2003}  \at{A model for
  combustion instability involving vortex shedding}.  \jt{Combust. Sci.
  Technol.}  \bvol{175}~(6),  \pg{1059--1083}.

\bibitem[Merk(1957)]{merk1957analysis}
{\sc \au{Merk, H.~J.}} \yr{1957}  \at{Analysis of heat-driven oscillations of
  gas flows}.  \jt{Appl. Sci. Res.}  \bvol{6}~(4),  \pg{317--336}.

\bibitem[Merk(1958)]{merk1958analysis}
{\sc \au{Merk, H.~J.}} \yr{1958}  \at{Analysis of heat-driven oscillations of
  gas flows}.  \jt{Appl. Sci. Res.}  \bvol{7}~(2),  \pg{175--191}.

\bibitem[Mondal {\em et~al.\/}(2017)Mondal, Unni \& Sujith]{mondal2017onset}
{\sc \au{Mondal, S.}, \au{Unni, V.~R.} \& \au{Sujith, R.~I.}} \yr{2017}
  \at{Onset of thermoacoustic instability in turbulent combustors: an emergence
  of synchronized periodicity through formation of chimera-like states}.
  \jt{J. Fluid Mech.}  \bvol{811},  \pg{659--681}.

\bibitem[Motter {\em et~al.\/}(2013)Motter, Myers, Anghel \&
  Nishikawa]{motter2013spontaneous}
{\sc \au{Motter, A.~E.}, \au{Myers, S.~A.}, \au{Anghel, M.} \& \au{Nishikawa,
  T.}} \yr{2013}  \at{Spontaneous synchrony in power-grid networks}.
  \jt{Nature Phys.}  \bvol{9}~(3),  \pg{191--197}.

\bibitem[Mukherjee {\em et~al.\/}(2015)Mukherjee, Heckl, Bigongiari, Vishnu,
  Pawar \& Sujith]{mukherjee2015nonlinear}
{\sc \au{Mukherjee, N.}, \au{Heckl, M.}, \au{Bigongiari, A.}, \au{Vishnu, R.},
  \au{Pawar, S.~A.} \& \au{Sujith, R.~I.}} \yr{2015} Nonlinear dynamics of a
  laminar {V} flame in a combustor.  \bt{In {\em Int. Congr. Sound Vib.\/}}, ,
  \vol{vol.~22}.

\bibitem[Nair \& Sujith(2014)]{nair2014multifractality}
{\sc \au{Nair, Vineeth} \& \au{Sujith, RI}} \yr{2014}  \at{Multifractality in
  combustion noise: predicting an impending combustion instability}.  \jt{J.
  Fluid Mech.}  \bvol{747},  \pg{635--655}.

\bibitem[Nair \& Sujith(2015)]{nair2015reduced}
{\sc \au{Nair, V.} \& \au{Sujith, R.~I.}} \yr{2015}  \at{A reduced-order model
  for the onset of combustion instability: {P}hysical mechanisms for
  intermittency and precursors}.  \jt{Proc. Combust. Inst.}  \bvol{35}~(3),
  \pg{3193--3200}.

\bibitem[Nair {\em et~al.\/}(2014)Nair, Thampi \&
  Sujith]{nair2014intermittency}
{\sc \au{Nair, V.}, \au{Thampi, G.} \& \au{Sujith, R.~I.}} \yr{2014}
  \at{Intermittency route to thermoacoustic instability in turbulent
  combustors}.  \jt{J. Fluid Mech.}  \bvol{756},  \pg{470--487}.

\bibitem[Nicoud {\em et~al.\/}(2007)Nicoud, Benoit, Sensiau \&
  Poinsot]{nicoud2007acoustic}
{\sc \au{Nicoud, F.}, \au{Benoit, L.}, \au{Sensiau, C.} \& \au{Poinsot, T.}}
  \yr{2007}  \at{Acoustic modes in combustors with complex impedances and
  multidimensional active flames}.  \jt{AIAA J.}  \bvol{45}~(2),
  \pg{426--441}.

\bibitem[Nicoud \& Wieczorek(2009)]{nicoud2009zero}
{\sc \au{Nicoud, F.} \& \au{Wieczorek, K.}} \yr{2009}  \at{About the zero mach
  number assumption in the calculation of thermoacoustic instabilities}.
  \jt{Int. J. Spray Combust. Dyn.}  \bvol{1}~(1),  \pg{67--111}.

\bibitem[Noiray(2017)]{noiray2017linear}
{\sc \au{Noiray, N.}} \yr{2017}  \at{Linear growth rate estimation from
  dynamics and statistics of acoustic signal envelope in turbulent combustors}.
   \jt{Trans. ASME J. Engng Gas Turbines Power}  \bvol{139}~(4),  \pg{041503}.

\bibitem[Noiray {\em et~al.\/}(2011)Noiray, Bothien \&
  Schuermans]{noiray2011investigation}
{\sc \au{Noiray, N.}, \au{Bothien, M.} \& \au{Schuermans, B.}} \yr{2011}
  \at{Investigation of azimuthal staging concepts in annular gas turbines}.
  \jt{Combust. Theory Model}  \bvol{15}~(5),  \pg{585--606}.

\bibitem[Noiray {\em et~al.\/}(2008)Noiray, Durox, Schuller \&
  Candel]{noiray2008unified}
{\sc \au{Noiray, N.}, \au{Durox, D.}, \au{Schuller, T.} \& \au{Candel, S.~M.}}
  \yr{2008}  \at{A unified framework for nonlinear combustion instability
  analysis based on the flame describing function}.  \jt{J. Fluid Mech.}
  \bvol{615},  \pg{139--167}.

\bibitem[Noiray \& Schuermans(2013)]{noiray2013deterministic}
{\sc \au{Noiray, N.} \& \au{Schuermans, B.}} \yr{2013}  \at{Deterministic
  quantities characterizing noise driven {H}opf bifurcations in gas turbine
  combustors}.  \jt{Int. J. Non-Linear Mech.}  \bvol{50},  \pg{152--163}.

\bibitem[{d}e Oliveira \& Abud(2020)]{de2020nonmonotonic}
{\sc \au{{d}e Oliveira, J.~F.} \& \au{Abud, C.~V.}} \yr{2020}  \at{Nonmonotonic
  critical threshold in the {K}uramoto model}.  \jt{Commun. Nonlinear Sci.
  Numer. Simul.}  \bvol{91},  \pg{105428}.

\bibitem[Pawar {\em et~al.\/}(2019)Pawar, Mondal, George \&
  Sujith]{pawar2019temporal}
{\sc \au{Pawar, S.~A.}, \au{Mondal, S.}, \au{George, N.~B.} \& \au{Sujith,
  R.~I.}} \yr{2019}  \at{Temporal and spatiotemporal analyses of
  synchronization transition in a swirl-stabilized combustor}.  \jt{AIAA J.}
  \bvol{57}~(2),  \pg{836--847}.

\bibitem[Paz{\'o}(2005)]{pazo2005thermodynamic}
{\sc \au{Paz{\'o}, D.}} \yr{2005}  \at{Thermodynamic limit of the first-order
  phase transition in the {K}uramoto model}.  \jt{Phys. Rev. E}  \bvol{72}~(4),
   \pg{046211}.

\bibitem[Pietras {\em et~al.\/}(2018)Pietras, Deschle \&
  Daffertshofer]{pietras2018first}
{\sc \au{Pietras, B.}, \au{Deschle, N.} \& \au{Daffertshofer, A.}} \yr{2018}
  \at{First-order phase transitions in the {K}uramoto model with compact
  bimodal frequency distributions}.  \jt{Phys. Rev. E}  \bvol{98}~(6),
  \pg{062219}.

\bibitem[Polifke(2014)]{polifke2014black}
{\sc \au{Polifke, Wolfgang}} \yr{2014}  \at{Black-box system identification for
  reduced order model construction}.  \jt{Ann. Nucl. Energy}  \bvol{67},
  \pg{109--128}.

\bibitem[Polifke(2020)]{polifke2020modeling}
{\sc \au{Polifke, W.}} \yr{2020}  \at{Modeling and analysis of premixed flame
  dynamics by means of distributed time delays}.  \jt{Prog. Energy Combust.}
  \bvol{79},  \pg{100845}.

\bibitem[Popovych {\em et~al.\/}(2005)Popovych, Maistrenko \&
  Tass]{popovych2005phase}
{\sc \au{Popovych, O.~V.}, \au{Maistrenko, Y.~L.} \& \au{Tass, P.~A.}}
  \yr{2005}  \at{Phase chaos in coupled oscillators}.  \jt{Phys. Rev. E}
  \bvol{71}~(6),  \pg{065201}.

\bibitem[Putnam(1971)]{putnam1971combustion}
{\sc \au{Putnam, A.~A.}} \yr{1971} {\em Combustion-driven oscillations in
  industry\/}.  \publ{Elsevier Publishing Company}.

\bibitem[Ramanan {\em et~al.\/}(2022)Ramanan, Ramankutty, Sreedeep \&
  Chakravarthy]{ramanan2022dynamical}
{\sc \au{Ramanan, V.}, \au{Ramankutty, A.}, \au{Sreedeep, S.} \&
  \au{Chakravarthy, S.~R.}} \yr{2022}  \at{Dynamical states of thermo-acoustic
  system with respect to frequency-phase relationship based on probabilistic
  oscillator model}.  \jt{Nonlinear Dyn.}  \pg{pp. 1--17}.

\bibitem[Ram{\'\i}rez-{\'A}vila {\em et~al.\/}(2018)Ram{\'\i}rez-{\'A}vila,
  Kurths \& Deneubourg]{ramirez2018fireflies}
{\sc \au{Ram{\'\i}rez-{\'A}vila, G.M.}, \au{Kurths, J.} \& \au{Deneubourg,
  J.~L.}} \yr{2018}  \at{Fireflies: A paradigm in synchronization}.  \bt{In
  {\em Chaotic, Fractional, and Complex Dynamics: New Insights and
  Perspectives\/}},  \pg{pp. 35--64}.  \publ{Springer}.

\bibitem[Rayleigh(1878)]{rayleigh1878explanation}
{\sc \au{Rayleigh, L.}} \yr{1878}  \at{The explanation of certain acoustical
  phenomena}.  \jt{Roy. Inst. Proc.}  \bvol{8},  \pg{536--542}.

\bibitem[Roy {\em et~al.\/}(2021)Roy, Singh, Nair, Chaudhuri \&
  Sujith]{roy2021flame}
{\sc \au{Roy, A.}, \au{Singh, S.}, \au{Nair, A.}, \au{Chaudhuri, S.} \&
  \au{Sujith, R.~I.}} \yr{2021}  \at{Flame dynamics during intermittency and
  secondary bifurcation to longitudinal thermoacoustic instability in a
  swirl-stabilized annular combustor}.  \jt{Proc. Combust. Inst.}
  \bvol{38}~(4),  \pg{6221--6230}.

\bibitem[Schuller {\em et~al.\/}(2020)Schuller, Poinsot \&
  Candel]{schuller2020dynamics}
{\sc \au{Schuller, T.}, \au{Poinsot, T.} \& \au{Candel, S.~M.}} \yr{2020}
  \at{Dynamics and control of premixed combustion systems based on flame
  transfer and describing functions}.  \jt{J. Fluid Mech.}  \bvol{894}.

\bibitem[Sethares(2007)]{sethares2007rhythm}
{\sc \au{Sethares, W.~A.}} \yr{2007} {\em Rhythm and transforms\/}.
  \publ{Springer Science \& Business Media}.

\bibitem[Shraiman(1986)]{shraiman1986order}
{\sc \au{Shraiman, B.~I.}} \yr{1986}  \at{Order, disorder, and phase
  turbulence}.  \jt{Phys. Rev. Lett.}  \bvol{57}~(3),  \pg{325}.

\bibitem[Shraiman {\em et~al.\/}(1992)Shraiman, Pumir, van Saarloos, Hohenberg,
  Chat{\'e} \& Holen]{shraiman1992spatiotemporal}
{\sc \au{Shraiman, B.~I.}, \au{Pumir, A.}, \au{van Saarloos, W.},
  \au{Hohenberg, P.~C.}, \au{Chat{\'e}, H.} \& \au{Holen, M.}} \yr{1992}
  \at{Spatiotemporal chaos in the one-dimensional complex {G}inzburg-{L}andau
  equation}.  \jt{Phys. D: Nonlinear Phenom.}  \bvol{57}~(3-4),  \pg{241--248}.

\bibitem[Singh {\em et~al.\/}(2022)Singh, Dutta, Dhadphale, Roy, Chaudhuri \&
  Sujith]{singh2022mean}
{\sc \au{Singh, S.}, \au{Dutta, A.~K.}, \au{Dhadphale, J.~M.}, \au{Roy, A.},
  \au{Chaudhuri, S.} \& \au{Sujith, R.~I.}} \yr{2022}  \at{Mean-field
  synchronization model for open-loop, swirl controlled thermoacoustic system}.
   \jt{arXiv preprint arXiv:2201.01764} .

\bibitem[Singh {\em et~al.\/}(2021)Singh, Roy, K.~V., Nair, Chaudhuri \&
  Sujith]{singh2021intermittency}
{\sc \au{Singh, S.}, \au{Roy, A.}, \au{K.~V., Reeja}, \au{Nair, A.},
  \au{Chaudhuri, S.} \& \au{Sujith, R.~I.}} \yr{2021}  \at{Intermittency,
  secondary bifurcation and mixed-mode oscillations in a swirl-stabilized
  annular combustor: Experiments and modeling}.  \jt{Trans. ASME J. Engng Gas
  Turbines Power}  \bvol{143}~(5),  \pg{051028}.

\bibitem[Strogatz(2000)]{strogatz2000kuramoto}
{\sc \au{Strogatz, S.~H.}} \yr{2000}  \at{From {K}uramoto to {C}rawford:
  exploring the onset of synchronization in populations of coupled
  oscillators}.  \jt{Phys. D: Nonlinear Phenom.}  \bvol{143}~(1-4),
  \pg{1--20}.

\bibitem[Strogatz(2018)]{strogatz2018nonlinear}
{\sc \au{Strogatz, S.~H.}} \yr{2018} {\em Nonlinear dynamics and chaos: with
  applications to physics, biology, chemistry, and engineering\/}.  \publ{CRC
  press}.

\bibitem[Strogatz {\em et~al.\/}(2005)Strogatz, Abrams, McRobie, Eckhardt \&
  Ott]{strogatz2005crowd}
{\sc \au{Strogatz, S.~H.}, \au{Abrams, D.~M.}, \au{McRobie, A.}, \au{Eckhardt,
  B.} \& \au{Ott, E.}} \yr{2005}  \at{Crowd synchrony on the {M}illennium
  {B}ridge}.  \jt{Nature}  \bvol{438}~(7064),  \pg{43--44}.

\bibitem[Subramanian {\em et~al.\/}(2013)Subramanian, Sujith \&
  Wahi]{subramanian2013subcritical}
{\sc \au{Subramanian, P.}, \au{Sujith, R.~I.} \& \au{Wahi, P.}} \yr{2013}
  \at{Subcritical bifurcation and bistability in thermoacoustic systems}.
  \jt{J. Fluid Mech.}  \bvol{715},  \pg{210--238}.

\bibitem[Sujith \& Pawar(2021)]{sujith2021thermoacoustic}
{\sc \au{Sujith, R.~I.} \& \au{Pawar, S.~A.}} \yr{2021}  \bt{
  \at{Thermoacoustic instability: A complex systems perspective}}.
  \publ{Springer Nature}.

\bibitem[Terada {\em et~al.\/}(2017)Terada, Ito, Aoyagi \&
  Yamaguchi]{terada2017nonstandard}
{\sc \au{Terada, Y.}, \au{Ito, K.}, \au{Aoyagi, T.} \& \au{Yamaguchi, Y.~Y.}}
  \yr{2017}  \at{Nonstandard transitions in the {K}uramoto model: a role of
  asymmetry in natural frequency distributions}.  \jt{J. Stat. Mech.}
  \bvol{2017}~(1),  \pg{013403}.

\bibitem[Wang {\em et~al.\/}(2021)Wang, Han, Song, Yang \& Sung]{wang2021multi}
{\sc \au{Wang, X.}, \au{Han, X.}, \au{Song, H.}, \au{Yang, D.} \& \au{Sung,
  C.~J.}} \yr{2021}  \at{Multi-bifurcation behaviors of stability regimes in a
  centrally staged swirl burner}.  \jt{Phys. Fluids}  \bvol{33}~(9),
  \pg{095121}.

\bibitem[Weng {\em et~al.\/}(2022)Weng, Unni, Sujith \&
  Saha]{weng2022synchronization}
{\sc \au{Weng, Yue}, \au{Unni, Vishnu~R}, \au{Sujith, RI} \& \au{Saha,
  Abhishek}} \yr{2022}  \at{Synchronization based model for turbulent
  thermoacoustic systems}.  \jt{arXiv preprint arXiv:2206.05663} .

\bibitem[Yu {\em et~al.\/}(2019)Yu, Jaravel, Ihme, Juniper \&
  Magri]{yu2019data}
{\sc \au{Yu, H.}, \au{Jaravel, T.}, \au{Ihme, M.}, \au{Juniper, M.~P.} \&
  \au{Magri, L.}} \yr{2019}  \at{Data assimilation and optimal calibration in
  nonlinear models of flame dynamics}.  \jt{Trans. ASME J. Engng Gas Turbines
  Power}  \bvol{141}~(12),  \pg{121010}.

\bibitem[Yu {\em et~al.\/}(2021)Yu, Juniper \& Magri]{yu2021data}
{\sc \au{Yu, H.}, \au{Juniper, M.~P.} \& \au{Magri, L.}} \yr{2021}  \at{A
  data-driven kinematic model of a ducted premixed flame}.  \jt{Proc. Combust.
  Inst.}  \bvol{38}~(4),  \pg{6231--6239}.

\bibitem[Zhang {\em et~al.\/}(2020)Zhang, Boccaletti, Liu \&
  Guan]{zhang2020synchronization}
{\sc \au{Zhang, J.}, \au{Boccaletti, S.}, \au{Liu, Z.} \& \au{Guan, S.}}
  \yr{2020}  \at{Synchronization of phase oscillators under asymmetric and
  bimodal distributions of natural frequencies}.  \jt{Chaos Solit. Fractals}
  \bvol{136},  \pg{109777}.

\bibitem[Zheng \& Zhang(2017)]{zheng2017modeling}
{\sc \au{Zheng, L.} \& \au{Zhang, X.}} \yr{2017} {\em Modeling and analysis of
  modern fluid problems\/}.  \publ{Academic Press}.

\bibitem[Zhou {\em et~al.\/}(2015)Zhou, Chen, Bi, Hu, Liu \&
  Guan]{zhou2015explosive}
{\sc \au{Zhou, W.}, \au{Chen, L.}, \au{Bi, H.}, \au{Hu, X.}, \au{Liu, Z.} \&
  \au{Guan, S.}} \yr{2015}  \at{Explosive synchronization with asymmetric
  frequency distribution}.  \jt{Phys. Rev. E}  \bvol{92}~(1),  \pg{012812}.

\end{thebibliography}


\begin{thebibliography}{1}
\expandafter\ifx\csname natexlab\endcsname\relax\def\natexlab#1{#1}\fi
\def\au#1{#1} \def\ed#1{#1} \def\yr#1{#1}\def\at#1{#1}\def\jt#1{\textit{#1}}
  \def\bt#1{#1}\def\bvol#1{\textbf{#1}} \def\vol#1{#1} \def\pg#1{#1}
  \def\publ#1{#1}\def\arxiv#1{#1}\def\org#1{#1}\def\st#1{\textit{#1}}

\bibitem[Balanov {\em et~al.\/}(2008)Balanov, Janson, Postnov \&
  Sosnovtseva]{balanov2008synchronization}
{\sc \au{Balanov, A.}, \au{Janson, N.}, \au{Postnov, D.} \& \au{Sosnovtseva,
  O.}} \yr{2008} {\em Synchronization: {F}rom simple to complex\/}.
  \publ{Springer Science \& Business Media}.

\end{thebibliography}

\end{document}

\maketitle
Here we consider the method of averaging on our model, derivation for recasting truncated equations from polar to Descartes coordinates, and estimated model parameters during the representative dynamical states in three turbulent combustors. Finally, we write the description of the movies S1 to S8 associated with the two different synchronization transitions.

\section{Method of averaging applied on the thermoacoustic mean-field model}\label{appA}
In this section, we apply the method of averaging on our thermoacoustic mean-field model. We differentiate $\eta (t)$ expressed in (\textcolor{blue}{3.10}) (shown in the manuscript) with respect to time ($t$) and obtain:
	\begin{equation}
	\dot{\eta} (t) = -\dot{R}(t)\cos\left[t + \Phi (t)\right] + R(t) \sin\left[t + \Phi (t)\right] + R(t)\dot{\Phi} (t) \sin\left[t + \Phi (t)\right].
	\label{eqa1}
	\end{equation}
	By representing the solution in the form of (\textcolor{blue}{3.10}) (shown in the manuscript), we express $\eta(t)$ as a function of $R(t)$ and $\Phi(t)$, which introduces additional ambiguity in the equation. To remove the introduced ambiguity, we have to prescribe one arbitrary relationship between these quantities, which we take as \citep{balanov2008synchronization}:
	\begin{equation}
	- \dot{R}(t) \cos\left[t + \Phi (t)\right] + R(t) \dot{\Phi} (t) \sin\left[t + \Phi (t)\right] = 0.
	\label{eqa2}
	\end{equation}
	Therefore, the derivative of $\eta (t)$ is a simple expression of the form:
	\begin{equation}
	\dot{\eta} (t) =  R(t) \sin\left[t + \Phi (t)\right].
	\label{eqa3}
	\end{equation}
	We follow \cite{balanov2008synchronization} for expressing $\eta (t)$, $\dot{\eta} (t)$ and $\ddot{\eta}(t)$ in terms of exponents of complex arguments. We start by reformulating the solution for $\eta (t)$ as:
	\begin{equation}
	\eta (t) = - \frac{1}{2 } \big(R(t) e^{i \Phi (t)} e^{it}  + R(t) e^{-i \Phi (t)} e^{-it}  \big) =  - \frac{1}{2}\big( a e^{it} + a^\ast e^{-it} \big).
	\label{eqa4}
	\end{equation}
	We introduce a complex function of time $a$, such that $a = R(t) e^{i \Phi (t)}$ and $a^\ast = R(t) e^{- i \Phi(t)}$, where the asterisk denotes the complex conjugate. Next, reformulating $\dot{\eta} (t)$, we get:
	\begin{equation}
	\dot{\eta} (t) = \frac{1}{2i} R(t) e^{i(t + \Phi (t))} - e^{-i(t + \Phi (t))}= - \frac{i}{2} \big(a e^{i t} - a^\ast e^{-i t} \big).
	\label{eqa5}
	\end{equation}
	Finally, reformulating $\ddot{\eta} (t)$ in terms of exponent of complex arguments, we obtain:
	\begin{equation}
	\ddot{\eta} (t) = -\frac{i}{2} \big( \dot{a}e^{i t} - \dot{a}^\ast e^{-i t} \big) + \frac{1}{2} \big(a e^{i t} + a e^{-i t}\big).
	\label{eqa6}
	\end{equation}
	We now substitute $\eta(t)$, $\dot{\eta}(t)$ and $\ddot{\eta}(t)$ ( \eqref{eqa4}, \eqref{eqa5} and \eqref{eqa6}, respectively) into (\textcolor{blue}{3.10}a) (shown in the manuscript) and we get:
	\begin{equation}
	\begin{split}
	 -i\big( \dot{a}e^{i t} & - \dot{a}^\ast e^{-i t} \big) +  \big(a e^{i t} + a e^{-i t}\big)   - i \zeta \big(a e^{i t} - a^\ast e^{-i t} \big) \\ & - \big( a e^{it} + a^\ast e^{-it} \big)  = -i\beta \cos(kz_f) \sum_{i=1}^N \big(e^{i(t + \theta_i (t))} - e^{-i(t + \theta_i (t))}\big).
	\label{eqa7}
	\end{split}
	\end{equation}
	By canceling the second and fourth terms in the above equation and then multiplying the whole equation by $e^{-i t}$, we obtain:
	\begin{equation}
	\begin{split}
	\big(\dot{a}  - \dot{a}^\ast e^{-2i t} \big)  +  \zeta \big(a - a^\ast e^{-2 i t} \big) =  \beta \cos(kz_f) \sum_{i=1}^N \big(e^{i\theta_i(t)} - e^{-2it} e^{-i \theta_i (t))}\big).
	\label{eqa8}
	\end{split}
	\end{equation}
	We now note that $a$, $\dot{a}$ and their complex conjugates are slow functions of time as compared to the functions $e^{\pm n i t}$, where $n$ is an integer. This means that the slow flow variables do not change much during one period of fast oscillations. If we average the whole equation over one period of fast oscillations, i.e., $T = 2 \pi$, we can eliminate the fast terms, and only the slow terms remain. The time average $\bar{f}$ of a smooth function $f(t)$ over the time interval $T$ is defined as $\bar{f} = \frac{1}{T} \int_0^{2 \pi} f(t) dt$. It is easy to see that all the terms containing $e^{-2i t}$ would integrate to zero over the time oscillation period $T$. Therefore, we apply time averaging on \eqref{eqa8} and obtain:
	\begin{equation}
	\dot{a} + \zeta a = \beta \cos(kz_f) \sum_{i=1}^N e^{i\theta_i(t)}.
	\label{eqa9}
	\end{equation}
	Recalling that $a = R(t) e^{i \Phi (t)}$ and substituting it in the above equation, we get:
	\begin{equation}
	\dot{R}(t) e^{i \Phi(t)} + i R(t) \dot{\Phi}(t) e^{i\Phi(t)} + \zeta R(t) e^{i \Phi (t)} =  \beta \cos(kz_f) \sum_{i=1}^N e^{i\theta_i(t)}.
	\label{eqa10}
	\end{equation}
	
	The above equation is further simplified in $\S$ \textcolor{blue}{3.4} of the manuscript for obtaining the amplitude and phase evolution equations.

\section{Truncated equations in Descartes coordinates}\label{appB}
The truncated equations (\textcolor{blue}{3.13a}) and (\textcolor{blue}{3.13b}) (shown in the manuscript) seems compact but their analysis is quite involved. Therefore, we recast the truncated equations from the polar coordinate to Descartes coordinate system. The time derivatives of the new variables ($A$ and $B$) can be expressed through $R$ and $\Phi$ as:
	\begin{subequations} 	\label{eqa11}
\begin{align}
	\dot{A}(t) & = \dot{R}(t)\cos\Phi(t) - R(t)\dot{\Phi}(t)\sin\Phi(t), \label{eqa11a} \\ 	
    \dot{B}(t) &= \dot{R}(t)\sin\Phi(t) + R(t)\dot{\Phi}\cos\Phi(t). \label{eqa11b}
\end{align}
\end{subequations}
On substituting \eqref{eqa11} into (\textcolor{blue}{3.13a}) and (\textcolor{blue}{3.13b}) (shown in the manuscript), we obtain the truncated equations in Descartes coordinates:
	\begin{subequations} 	\label{eqa12}
\begin{align}
	\dot{A}(t) &= \beta \cos(k z_f) \sum_{i=1}^{N} \left[\cos\left(\theta_i(t) - \Phi(t)\right) \cos\Phi(t) - \sin\left(\theta_i(t) - \Phi(t)\right) \sin\Phi(t)  \right] - \zeta A(t), \label{eqa12a} \\ 	
    \dot{B}(t) &= \beta \cos(k z_f) \sum_{i=1}^{N} \left[\cos\left(\theta_i(t) - \Phi(t)\right) \sin\Phi(t) - \sin\left(\theta_i(t) - \Phi(t)\right) \cos\Phi(t)  \right] - \zeta B(t), \label{eqa12b}
\end{align}
\end{subequations}

\section{Estimated model parameters corresponding to dynamical states in experiments}
\label{appC}

The estimated model parameters are shown in the tables \ref{tab1}, \ref{tab2} and \ref{tab3} corresponding to the bluff-body stabilized, swirl-stabilized, and annular combustor, respectively. The damping coefficient ($\zeta$) is
obtained using parameter optimisation \textcolor{blue}{(4.5)} and is subsequently fixed for determining other states during the transition. The value of $\zeta$ during the state of combustion noise is obtained as 0.6, 0.4, and 0.3 for the bluff-body stabilized, swirl-stabilized, and annular combustor, respectively. 

 \begin{center} 
   \captionof{table}{Parameter estimated for different dynamical states observed in the bluff-body stabilized combustor.}
   \begin{tabular}{ c c c c c c c}     
   \hline
   States & $K$ & $\eta$ & $\dot{\eta}$ & $\theta_m$ & $\sigma$ & $L$ \\ \hline
   Combustion noise & 0.230 & 0.011 & 0.001 & 0.499 & 0.100 & 0.004 \\  \hline
     Intermittency & 0.752 & 0.015 & 0.009 & 0.497 & 0.101 & 0.009 \\  \hline
    LCO & 1.997 & 0.009 & 0.007 & 0.499 & 0.100 & 0.900 \\  \hline
    \label{tab1}
   \end{tabular}
   \end{center}
 
   \begin{center} 
   \captionof{table}{Parameter estimated for different states observed in the swirl-stabilized combustor.}
   \begin{tabular}{ c c c c c c c}     
   \hline
   States & $K$ & $\eta$ & $\dot{\eta}$ & $\theta_m$ & $\sigma$ & $L$ \\ \hline
   Combustion noise & 0.506 & 0.108 & -0.003 & 0.368 & 0.150 & 0.005 \\  \hline
    Intermittency & 0.920 & 0.098 & -0.009 & 0.366 & 0.130 & 0.055 \\  \hline
    Low-amplitude LCO & 1.240 & -0.040 & -0.030 & 0.500 & 0.098 & 0.130 \\  \hline
    High-amplitude LCO & 1.542 & -0.116 & 0.180 & 1.390 & 0.094 & 1.100 \\  \hline
    \label{tab2}
   \end{tabular}
   \end{center}
   
   \begin{center} 
   \captionof{table}{Parameter estimated for different states observed in the annular combustor.}
   \begin{tabular}{ c c c c c c c}     
   \hline
   States & $K$ & $\eta$ & $\dot{\eta}$ & $\theta_m$ & $\sigma$ & $L$ \\ \hline
   Combustion noise & 1.091 & 0.085 & -0.015 & 0.428 & 0.121 & 0.003 \\  \hline
    Intermittency & 1.367 & 0.036 & -0.001 & 0.494 & 0.101 & 0.005 \\  \hline
    Low-amplitude LCO & 1.550 & 0.022 & 0.009 & 0.498 & 0.101 & 0.040 \\  \hline
    High-amplitude LCO & 1.710 & -0.258 & 0.689 & 0.199 & -0.302 & 0.319 \\  \hline
    \label{tab3}
   \end{tabular}
   \end{center}

\section{Movies}

Movies S1-S4 feature the emergence of global synchronization among the oscillators during the \textit{second-order} continuous synchronization for the bluff-body stabilized combustor, while movies S5-S8 depict the emergence of synchronization during the \textit{first-order} explosive synchronization for the annular combustor. \\

\textbf{Movie S1}: Panel (a) shows the instantaneous relative phase ($\psi_i$) between the phase of the heat release rate fluctuations ($\theta_i$) and acoustic pressure ($\Phi$) obtained from experiments in the bluff-body stabilized combustor during the occurrence of combustion noise ($\phi=0.86, K=0.23$). The flow goes from left to right. Also shown is the probability distribution of $\psi_i$. In panel (b), the instantaneous oscillator distribution is shown in the $\dot{\theta}_i - \dot{\psi}_i$ phase space along with the probability distribution for instantaneous frequencies ($\dot{\theta}_i$) both from the spatiotemporal measurements (light shade) and the model (dark shade). Here, $\dot{\theta}_i$ is normalized, and the mean acoustic frequency $\Omega_0$ has been subtracted. In panel (c), the instantaneous distribution of oscillators ($\theta_i$) is shown in polar coordinates. The frame of reference of the oscillators is co-rotating with respect to acoustic frequency $\Omega_0$, and the phase of acoustic pressure is indicated by the dashed line. The movie shows the desynchronized behavior of the phase-field during the stable operation of the combustor. The oscillators have a broadband distribution of ($\dot{\theta}_i$) and ($\psi_i$). Phasors have been colored as blue if $|\psi_i| < \pi/2$ and red otherwise to delineate regions of acoustic power sources and sinks. Note in particular the presence of many peaks in the frequency distribution close to $\dot{\theta}_i=0$ in panel (b).\\

\textbf{Movie S2}: All the panels are the same as in Movie S1. This movie exhibits the behavior during the periodic part of intermittency ($\phi=0.72, K=0.75$). The behavior of the oscillators during the aperiodic part of intermittency is similar to Movie S1. Compared to movie S1, notice the larger regions of phase-synchronized clusters and narrower distributions of $\mathcal{P}(\psi_i)$ and $\mathcal{P}(\dot{\theta}_i)$. This state is referred to as chimera, where regions of synchronization and desynchronization co-exists. We also observe that the frequency distribution $\mathcal{P}(\dot{\theta}_i)$ is clearly unimodal, implying that the frequency clusters in Movie S1 have started synchronizing to the mean acoustic frequency.\\

\textbf{Movie S3}: All the panels are the same as in Movie S1. This movie illustrates the behavior of oscillators at $\phi=0.65$ and $K=1.25$ where limit cycle behavior starts to emerge. In the movie, the size of the phase-synchronized cluster is clearly much larger than that observed in Movie S1 and S2. The oscillators also display narrowband distribution of relative phases ($\psi_i$), implying a much large degree of phase synchronization. The distribution of instantaneous frequencies ($\dot{\theta}_i$) is clearly unimodal and possesses a strong peak near the acoustic frequency, indicating the entrainment of more and more oscillators. \\

\textbf{Movie S4}: All the panels are the same as in Movie S1. This movie shows the behavior of oscillators during thermoacoustic instability at $\phi=0.56$ and $K=2$. We notice the phase-field inside the combustor is coherent, and all the oscillators exhibit global phase synchronization. All the oscillators are entrained at the mean acoustic frequency and phase-locked with the distribution $\psi_i < |\pi/2|$, implying global phase synchronization among the oscillators. We also notice a very sharp, narrowband distribution in the distribution $\mathcal{P}(\dot{\theta}_i)$.\\

Movie S1 to S4 clearly shows the gradual emergence of a single peak in the frequency distribution $\mathcal{P}(\dot{\theta}_i)$ near the acoustic frequency $\Omega_0$. This gradual emergence is associated with continuous growth in the size of the cluster, leading to a continuous \textit{second-order} synchronization transitions.\\

\textbf{Movie S5}: All the panels are the same as in Movie S1. This movie shows the oscillator behavior in the annular combustor for 4 representative burners during the state of combustion noise ($\phi=0.44, K=1.09$). We can clearly observe the desynchronized behavior of the phase-field. The oscillators have broadband phase ($\dot{\theta}_i$) and instantaneous frequency ($\psi_i$) distribution. Notice that $\mathcal{P}(\dot{\theta}_i)$ has a broad peak at $\dot{\theta}_i\lesssim -0.5$ in panel (b).\\

\textbf{Movie S6}: All the panels are the same as in movie S1. The movie shows oscillator behavior in the annular combustor during the periodic part of intermittency ($\phi=0.47, K=1.37$). In this state, the oscillators display synchronous and asynchronous regions, as observed in chimeras. The distribution $\mathcal{P}(\dot{\theta}_i)$ shows a slightly narrower peak than that shown in Movie S5.\\

\textbf{Movie S7:} All the panels are the same as in movie S1. The movie shows oscillator behavior in the annular combustor during low-amplitude thermoacoustic instability ($\phi=0.49, K=1.55$). The distribution of $\mathcal{P}(\psi_i)$ is much narrower with a significant number of oscillators lying within $|\psi_i|<\pi/2$. We also notice the appearance of a secondary peak in the distribution of $\mathcal{P}(\dot{\theta}_i)$. The peak implies the existence of presence of two synchronized clusters.\\

\textbf{Movie S8}: All the panels are the same as in movie S1. The movie shows oscillator behavior in the annular combustor during high-amplitude thermoacoustic instability ($\phi=0.52, K=1.71$). The movie shows global phase synchronization where all the oscillators are entrained at the frequency of acoustic fluctuations. The phase and frequency distributions are narrowband, and a majority of phase oscillators lie within $|\psi|<\pi/2$. \\

Movies S5-S8 show the manner in which the phase oscillators evolve to the state of global phase synchronization. Crucially, the appearance of bimodal distribution in Movie S7 implies the existence of two synchronized clusters, which coalesce abruptly when $K$ is increased beyond a critical value. Such an abrupt emergence of one single large cluster of synchronized oscillators leads to abrupt \textit{first-order} explosive synchronization.

\bibliographystyle{jfm}
\bibliography{references}